\documentclass[12pt,preprint]{aastex}

\newif\ifjournal\journalfalse

\usepackage{graphicx}
\usepackage{bm}
\renewcommand{\vec}[1]{\boldsymbol{#1}}

\newcommand{\grad}{\nabla}
\renewcommand{\div}{\nabla \cdot}
\newcommand{\rot}{\nabla \times}
\newcommand{\pp}[2]{\frac{\partial #1}{\partial #2}}
\def\jgr{{\itshape J. Geophys. Res.} }
\def\apj{{\itshape Astrophys. J.} }
\def\prl{{\itshape Phys. Rev. Lett.} }
\def\pop{{\itshape Phys. Plasmas} }
\def\aap{{\itshape Astron. Astrophys.} }

\def\mnras{{\itshape Monthly Notices of the RAS} }

\shorttitle{RELATIVISTIC TWO-FLUID RECONNECTION}
\shortauthors{Zenitani, Hesse, \& Klimas}

\begin{document}

\title{Two-Fluid Magnetohydrodynamic Simulations of Relativistic Magnetic Reconnection}

\author{Seiji Zenitani, Michael Hesse, and Alex Klimas}
\affil{
NASA Goddard Space Flight Center, Greenbelt, MD 20771;
Seiji.Zenitani-1@nasa.gov
}

\begin{abstract}
We investigate
the large-scale evolution of a relativistic magnetic reconnection
in an electron--positron pair plasma
by a relativistic two-fluid magnetohydrodynamic (MHD) code.
We introduce an interspecies friction force
as an effective resistivity to dissipate magnetic fields.
We demonstrate that
magnetic reconnection successfully occurs in our two-fluid system, and
that it involves Petschek-type bifurcated current layers in a later stage.
We further observe a quasi-steady evolution thanks to an open boundary condition,
and find that the Petschek-type structure is stable over the long time period.
Simulation results and theoretical analyses exhibit that
the Petschek outflow channel becomes narrower
when the reconnection inflow contains more magnetic energy,
as previously claimed.  Meanwhile, we find that
the reconnection rate goes up to $\sim$1 in extreme cases,
which is faster than previously thought.
The role of the resistivity,
implications for reconnection models in the magnetically dominated limit,
and relevance to kinetic reconnection works are discussed.
\end{abstract}


\section{INTRODUCTION}

Magnetic reconnection in collisionless or collisional plasmas
is the driver of explosive events in space and astroplasmas.
By breaking the magnetic field topology,
it rapidly releases the magnetic energy into plasma kinetic energy
in a short timescale, and therefore it explains
particle acceleration or bursty emission signatures in these sites. 
On the Sun, it is widely recognized that
magnetic reconnection drives solar flare or coronal mass ejections
(see \citet{aschwanden} for review).
Theoretical models have long been established \citep{sweet,parker,petschek}, and
a series of MHD simulations make a significant success
to understand flare-type events (e.g., \citet{chen00,yoko01}).

Magnetic reconnection is considered
in a wide variety of high-energy astrophysical contexts too.
For example, soft gamma repeaters (SGRs) and
anomalous X-ray pulsars (AXPs) are now best described by
a ``magnetar'' model \citep{duncan92,woods06},
a neutron star with superstrong magnetic fields up to $10^{14}-10^{15} G$.
In analogy to the Sun,
flares on and around the magnetar \citep{thom95,thom01,lyut03a,lyut06}
are considered as driving mechanism of bursty events,
in relativistic electron--positron environments.
Such flares, or magnetic reconnection events,
should be strongly influenced by the relativistic effects,
because the ultra strong magnetic field boosts
the Alfv\'{e}n speed up to the light speed. 

The pulsar environments are also influenced by relativistic plasmas and
the strong magnetic fields ($\sim 10^{12} G$) of the neutron star.
Recent time-dependent simulations of pulsar magnetospheres
\citep{kom06,buc06,spit06} suggested that
the magnetic reconnection near the $Y$ point,
where the outmost closed field lines intersect
the equatorial current sheet, is of critically importance,
while these models cannot deal with local reconnection physics.
Outside the magnetosphere,
reconnection processes in the ``striped'' current sheets
are considered to dissipate magnetic energy
inside the relativistic plasma outflow
(pulsar winds; \citet{michel82,michel94,coro90,lyu01,kirk03}) and
its termination shock \citep{lyu03}.
Furthermore, active galactic nuclei \citep{dimatteo,birk01},
extragalactic jets \citep{lb98},
gamma-ray burst (GRB) outflows \citep{dr02,drs02},
and potentially the black hole ergosphere \citep{koide08}
may be influenced by the magnetic reconnection in the relativistic regime.
Indeed, there is a high demand for modeling
the magnetic reconnection in these relativistic environments.

However, the relativistic theory of a magnetic reconnection is
not yet well established.
\citet{bf94b} extended the steady state reconnection models
into the relativistic regime,
based on a relativistic extension of Ohm's law \citep{bf93}.
Assuming uniform proper density,
they argued that the Lorentz boost may enhance the energy conversion rate
both in Sweet--Parker and in Petschek reconnections.
In the Sweet--Parker regime, \citet{lyut03b} further examined this idea and
claimed that reconnection outflow may be super-Alfv\`{e}nic.
On the other hand, \citet{lyu05} pointed out that
the reconnection will not be fast
because the relativistic gas pressure increases the outflow inertia.
Recently, the authors discussed a two-fluid description
and we showed that
the incompressibility assumption is invalid for relativistic outflow \citep{zeni08c}.
In the Petschek regime,
in which the reconnection involves a bifurcated slow-shock structure,
\citet{lyu05} argued that
the reconnection would not be an efficient energy converter
because the slow-shock angle becomes narrower.

Meanwhile, there has been a remarkable progress
on the kinetic-scale behaviors of relativistic magnetic reconnection,
by self-consistent particle-in-cell (PIC) simulations.
\citet{zeni01} demonstrated that
powerful DC acceleration occurs
around the reconnecting $X$-type region.
This and the relevant particle acceleration
generate nonthermal plasma distributions
on a larger scale \citep{claus04,zeni07,bessho07,karl08},
and particle acceleration may be enhanced
in a compressed pulsar--wind configuration \citep{lyu08}.
In the orthogonal plane, the current-driven drift kink instabilities
are of importance \citep{zeni05a,zeni07},
because they grow faster and may interfere with the magnetic reconnection.
Due to a wide variety of such plasma instabilities
the reconnection current sheet exhibits complex evolution
in three dimensions \citep{claus04,zeni05b,zeni08}.
Furthermore, it was recently pointed out
that kinetic effects are important
not only in the critical reconnecting region \citep{hesse07},
but also in the reconnection outflow region
as an anisotropy-driven Weibel-type instability \citep{zeni08b}.
However, these PIC simulations typically deal with
the spatial domain of several hundreds of
the plasma inertial length $(c/\omega_p)$ in two or three dimensions.
The large-scale evolution of relativistic reconnection systems
is still an open problem.

In order to study large-scale properties of
a relativistic magnetic reconnection
beyond these kinetic scales,
and
in order to investigate larger scale astrophysical problems
which contain relativistic magnetic reconnection
such as magnetar flares and global pulsar magnetospheres,
we need a relativistic extension of
magnetohydrodynamic (MHD) codes (see \citet{marti03} for review).
However, relativistic hydrodynamic codes are difficult to develop,
because of the complexity of the equation system.
In particular, these codes typically use
an inverse transformation
from the conserved variables in the lab frame
to the primitive variables in the proper frame.
This can be calculated
by solving quartic equations, or
by using iterative methods (e.g. \citet{duncan94}).
Such inverse conversion is further complicated
in the ideal MHD cases \citep{koide96,kom99,dz03,noble06}.
Overcoming these difficulties, there has been a remarkable progress
both in relativistic magnetohydrodynamic (RMHD) codes
and in general relativistic magnetohydrodynamic (GRMHD) codes
\citep{koide99,gammie03,mizuno06}.

To deal with the magnetic reconnection problems,
one has to incorporate ``resistive'' effects into the RMHD equations.
Otherwise, only the numerical resistivity plays
a role to dissipate magnetic fields. 
The first resistive RMHD work was done by \citet{naoyuki06},
by using a spatially limited resistivity.
Although their system size is very small (416 $\times$ 200),
they successfully presented
a Petschek-type reconnection in a mildly relativistic regime.
\citet{kom07} also developed
the upwind scheme for resistive RMHD,
which may be applicable to the reconnection problem.
These resistive RMHD studies are based on
a simple form of time-stationary Ohm's law \citep{bf93}.

In the present paper,
we investigate large-scale properties of
a relativistic magnetic reconnection
in an electron--positron pair plasma
by means of two-fluid RMHD simulations.
In contrast to the conventional RMHD models,
we introduce a relativistic two-fluid approximation
for the first time to our knowledge,
so that we can describe the physics in more detail.
An interspecies friction term is introduced
in the momentum equations,
which works as an effective resistivity.
By using a spatially limited resistivity profile,
we successfully reproduce a magnetic reconnection.
We also note that we carry out larger scale simulations,
directly solving equations to restore the primitive variables.

This paper is organized as follows.
In Section 2, we describe our simulation model.
Mathematical procedures are also presented in the appendix chapters.
In Section 3, we overview the system evolution in detail,
and present parameter dependences.
Especially, we analyze the structure of
bifurcated Petschek-type current layers in depth.
We also demonstrate that the system evolution
highly depends on the resistivity model.
In Section 4, we discuss
the characteristics of the two-fluid approach and
implications for the reconnection in the magnetically dominated limit.
The last section Section 5 contains the summary.

\section{SIMULATION MODEL}

We employ a relativistic two-fluid model of electrons and positrons.
The electron motion and positron motion are considered separately.
The continuity equation, the momentum equation, and
the energy equation of relativistic positron fluid,
and Maxwell equations are as follows.
In addition, we introduced an interspecies friction term to the momentum equation,
which is proportional to the relative motion of electrons and positrons.
\begin{eqnarray}
\pp{N_p}{t} &=& \pp{}{t} \gamma_p n_p
= -\div (n_p \vec{u}_p) \\
\label{eq:mom}
\pp{\vec{m}_p}{t} &=& \pp{}{t} \Big( \frac{ \gamma_p w_p \vec{u}_p }{c^2} \Big)
= -\div \Big( \frac{ w_p \vec{u}_p\vec{u}_p }{c^2} + \delta_{ij} p_p \Big) \nonumber \\
&&
+ \gamma_p n_p q_p (\vec{E}+\frac{\vec{v}_p}{c}\times\vec{B})
- \tau_{fr} N_p N_e (\vec{v}_p-\vec{v}_e) \\
\label{eq:ene}
\pp{K_p}{t} &=& \pp{}{t} \Big(\gamma_p^2 w_p - p_p - N_p m c^2 \Big) \nonumber \\
&=& -\div ( \gamma_p w_p \vec{u}_p - n_p m c^2\vec{u}_p ) + \gamma_p n_p q_p (\vec{v}_p\cdot\vec{E}) \\
\pp{\vec{B}}{t} &=& - c \rot \vec{E} \\
\pp{\vec{E}}{t} &=& c \rot \vec{B} - 4\pi \sum_{s=p,e} {q_s n_s \vec{u}_s}
\end{eqnarray}
In these equations,
the subscript $s$ denotes the species
(``$p$'' for positrons, and ``$e$'' for electrons),
$N$ is the lab-frame density,
$\gamma$ is the Lorentz factor,
$n$ is the proper density,
$\vec{u}=\gamma\vec{v}$ is the fluid 4-velocity,
$\vec{m}$ is the momentum density,
$w$ is the specific enthalpy,
$\delta_{ij}$ is the Kronecker delta,
$p$ is the proper isotropic pressure,
$q_p=-q_e$ is the positron/electron charge,
$\tau_{fr}$ is the coefficient for an inter-species friction, and
$K$ is the kinetic energy density
(energy density without the rest mass energy).
The enthalpy $w$ is defined in the following way:
\begin{eqnarray}
w=e+p=nmc^2+[\Gamma/(\Gamma-1)]p=hnmc^2,
\end{eqnarray}
where $e$ is the internal energy,
$\Gamma=4/3$ is the specific heat, and
$h$ is the dimensionless specific enthalpy.

We solve the equations by using modified Lax--Wendroff scheme.
To restore the primitive variables ($n, p, \gamma, \vec{u}$)
from the conservative variables $N$, $\vec{m}$, and $K$,
we use the following quartic relation for $\bar{u}=|\vec{u}|/c$
\begin{eqnarray}
\label{eq:fourth}
f(\bar{u}) &=& G^2(\mathcal{E}^2-M^2) \bar{u}^4 - 2GMD \bar{u}^3
\nonumber \\ &&
+ \Big[ G^2\mathcal{E}^2 - D^2 -2GM^2(G-1) \Big] \bar{u}^2 
\nonumber \\ &&
- 2DM(G-1) \bar{u} - (G-1)^2M^2 = 0
\end{eqnarray}
where $D = Nmc^2$, $M=|\vec{m}|c$, $\mathcal{E}=K+D$, and $G=\Gamma/(\Gamma-1)$.
We algebraically solve this equation by
decomposing the quartic equation into
the product of two quadratic equations.
See Appendices A and B for details.
We stop the simulation
when we find multiple possible solutions or
when the solution is physically invalid (e.g., negative density).
We added small artificial viscosity to the code,
which works when the fluid 4-velocity has a strong shear
so that it reduces a numerical oscillation near discontinuities.

We study the system evolution in the two-dimensional $x$-$z$ plane.
We choose the following relativistic Harris model
as an initial configuration:
\begin{eqnarray}
\vec{B} &=& B_0 \tanh(z/L)~\vec{\hat{x}} \\
\vec{E} &=& \eta_{eff} \vec{j} \label{eq:ini_E} \\
\vec{j} &=& 2 q_p n_0 u_0 \cosh^{-2}(z/L)~\vec{\hat{y}}
= \sum_{s=p,e} q_s n_s \vec{u}_{s}(z) \\
n_{s} &=& n_0 \cosh^{-2}(z/L) + n_{in} \\
p_{s} &=& p_0 \cosh^{-2}(z/L) + p_{in}
\end{eqnarray}
where $L$ is the typical half-thickness of the current sheet.
In the electric field,
$\eta_{eff}=(\tau_{fr}/q_p^2)$ is an effective resistivity,
and $u_0$ stands for the initial positron drift to carry the current.
We also consider uniform background plasmas
whose density and pressure are $n_{in}$ and $p_{in}$, respectively.
In this work, the plasma pressure in the Harris sheet
is set to $p_0=n_0mc^2$.
The background pressure is set to $p_{in}=n_{in}mc^2$
unless stated otherwise.

In the case of two-dimensional antiparallel reconnection,
we already know that positron motion and electron motion are
the same in the $x$-$z$ plane and the opposite in the $y$-direction.
Therefore, we assume the following symmetric motion
$u_{px}=u_{ex}$, $u_{py}=-u_{ey}$, $u_{pz}=u_{ez}$,
$n_{p}=n_{e}$, $p_{p}=p_{e}$
so that we can reduce the computational cost.
Consequently, the current has only the $y$-component and $j_x=j_z=0$,
and we can neglect three components of the electromagnetic field, $E_x=E_z=B_y=0$.
The assumption also justifies that
we do not consider the interspecies energy transfer in equation \ref{eq:ene},
because we assume such a symmetric model.
In PIC simulations,
one characteristic process to generate the charge separation
is the Weibel instability \citep{zeni08b},
driven by an anisotropy in plasma distribution function;
however, such a small-scale kinetic effect
is out of scope of this fluid paper.
In a MHD-scale, charge neutrality is plausible.
By assumption, we do not need to deal with
the Poisson equations $\div{\vec{E}} = 4\pi \sum_{s} (q_s \gamma_s n_s)$
in this system.

We introduce a spatially localized resistivity
by controlling the interspecies friction force.
Its profile is set in the following way:
\begin{eqnarray}
\label{eq:localized}
\tau_{fr} = \tau_0 + \tau_1 \cosh^{-2}[ {\sqrt{x^2+z^2}/(2L)} ],
\end{eqnarray}
where the background value $\tau_0$ is
equivalent to the Reynolds number $S=3000$,
and localized value $\tau_1$ is equivalent to $S=30$.
In addition, a magnetic field perturbation is added to the initial model
to quickly trigger a magnetic reconnection.
It is defined by the following vector potential:
\begin{equation}
\delta A_y = 2 L B_1 \exp[-(x^2+z^2)/(2L)^2],
\end{equation}
where $B_1=0.03B_0$ is the typical peak amplitude of the perturbed field.

The boundaries are located at $x=\pm L_x$ and $z=\pm L_z$,
and the reconnection is considered around the origin.
Boundary conditions for the fluid properties, the electric field, and
the tangential magnetic field are set to open:
$\partial / \partial x=0$ at the $x$-boundaries (outflow boundaries)
and
$\partial / \partial z=0$ at the $z$-boundaries (inflow boundaries).
The normal component of the magnetic field is set
so that it satisfies $\div \vec{B} = 0$ at the boundaries.
The system size ($2L_x \times 2L_z$) is presented
in Table \ref{table} in the unit of $L$.
The thickness is typically resolved by 20 grids ($L=20\Delta_g$).
This grid size is selected so that
it is comparable to the kinetic scale of
a typical gyroradius $\Delta_g \sim (mc^2/q_pB_0)=0.05L$
(Equations (15) and (16) in \citet{zeni07}).
In this equilibrium,
the electron inertia length is
$c/\omega_p = [ mc^2 /(4\pi \gamma_{\beta} n_0 q_p^2) ]^{1/2} \simeq u_z [ n_0mc^2/p_0 ]^{1/2} = 0.1$,
based on the reference density $n_0$.
The time step is set to $\Delta t=0.2(\Delta_g/c)=0.01~\tau_c$,
where $\tau_c = L/c$ is the light transit time.
It is sufficiently small,
$(q_pB_0/hmc) \Delta t \sim 0.04 \ll 2\pi$,
with respect to the fluid bulk motion.


Our code is originally developed from the CANS code,
a collection of hydrodynamic and MHD codes,
which has been extensively used
in Japanese solar and astrophysical community.
The code is massively parallelized by MPI.

We carry out various simulation runs with different parameters.
The list of simulation runs is presented in Table \ref{table}.
The parameter $\sigma_m$ is the magnetization parameter,
which stands for
the ratio of the magnetic energy flow to the rest mass energy flow,
\begin{eqnarray}
\label{eq:sigma1}
\sigma_m = \frac{B_0^2}{4\pi m (2\gamma^2n) c^2}.
\end{eqnarray}
Another parameter $\sigma_{\varepsilon}$ is
the exact ratio of the magnetic energy flow to the plasma energy flow,
which contains relativistic pressure effect
\begin{eqnarray}
\label{eq:sigma2}
\sigma_{\varepsilon} = \frac{B_0^2}{4\pi (2\gamma^2 w)}.
\end{eqnarray}
The Alfv\'{e}n speed $c_{A}$ in the relativistic regime
can be written as follows:
\begin{eqnarray}
c_{A}=\sqrt{\frac{\sigma_{\varepsilon}}{1+\sigma_{\varepsilon}}}.
\end{eqnarray}
The subscript $in$ ($\sigma_{m,in},\sigma_{\varepsilon,in}$
and $c_{A,in}$) stands for the upstream values,
based on the initial inflow properties (e.g. $n_{in},p_{in}$).
Later we often use $\sigma_{\varepsilon,in}$
as a measure of the upstream energy composition.
In Table \ref{table},
run U3 employs the uniform resistivity model
without the $\tau_1$ term in equation \ref{eq:localized}.
Runs S3, M3 and XL3 are done in different resolutions.

Before visiting the simulation results,
let us clarify the role of a newly introduced friction term.
From the positron momentum equation (Equation \ref{eq:mom}),
we obtain the following relation:
\begin{eqnarray*}
\vec{E}
&+& \frac{\vec{v}_p}{c}\times\vec{B} \\
&=& \frac{1}{\gamma_p n_p q_p}
\Big[
{ n_pm_p (\vec{u}_p \cdot \grad) h_p \vec{u}_p +
m_p h_p \vec{u}_p [ \div ( n_p \vec{u}_p) ] }
\\
&&
+ \grad p_p
+ \pp{}{t} \gamma_p m_p n_p h_p\vec{u}_p
\Big]  + \frac{\tau_{fr}N_pN_e}{\gamma_p n_p q_p}
(\vec{v}_{p}-\vec{v}_{e})\\
&=&
\frac{m_p}{q_p}
\Big( \pp{}{t} + \vec{v}_p \cdot \grad \Big) h_p\vec{u}_p
+
\frac{1}{\gamma_p n_p q_p}\grad p_p
+ \frac{\tau_{fr}N_pN_e}{\gamma_p n_p q_p}
(\vec{v}_{p}-\vec{v}_{e})
.
\end{eqnarray*}
We consider Ohm's law in the $y$ direction.
Dropping $\partial /\partial y$ and
considering symmetric electron properties,
we obtain
\begin{eqnarray}
\label{eq:ohm}
E_y + ( \frac{\vec{v}_p}{c}\times\vec{B} )_y
&=& \frac{m_p}{q_p}
\Big[
\pp{(h_pu_{py})}{t} + v_{px}\frac{\partial (h_pu_{py})}{\partial x}
\nonumber \\
& &+
{v}_{pz} \frac{\partial (h_pu_{py})}{\partial z} \Big]
+ \eta_{eff} j_y.
\end{eqnarray}
Thus, the fluid inertial effect, the momentum advection, and
the interspecies friction term work as an effective resistivity.
In our two-fluid model, this interspecies resistivity
plays an essential role to sustain the magnetic reconnection.
Around the reconnecting $X$-point,
the Lorentz term is negligible because $\vec{B}\sim 0$ and $v_x,v_z\sim 0$,
the advection terms usually vanish by symmetry,
and the inertial terms do not work
in the quasi-steady condition ($\partial /\partial t \sim 0$).
Therefore, the interspecies resistivity sustains
the reconnection electric field $E_y \sim \eta_{eff}j_y$.
Note that the reconnection cannot go on
without the reconnection electric field $E_y$.
We do not assume any specific mechanism as the interspecies friction term.
In a collisional regime,
it should be equivalent to the collisional term;
however, we do not know the true form of the relativistic collisional term,
which often relies on empirical functions (e.g., Section 7 in \citet{clare86}).
In a collisionless regime,
it is known that the off-diagonal part of the pressure tensor
sustains the reconnection electric field \citep{hesse07}
in the kinetic simulations.
Although its physical meaning is not yet well established,
the off-diagonal part of the pressure tensor
contains several kinetic effects
such as the escaping convection of the accelerating particles,
or the inertial effect of thermal plasma populations.
The purpose of the interspecies resistivity is
to represent these kinetic effects in the fluid approximation,
for the purpose of larger scale modeling.

\begin{figure}[htbp]
\begin{center}
\ifjournal
\includegraphics[width={0.85\columnwidth},clip]{f1.eps}
\else
\includegraphics[width={0.85\columnwidth},clip]{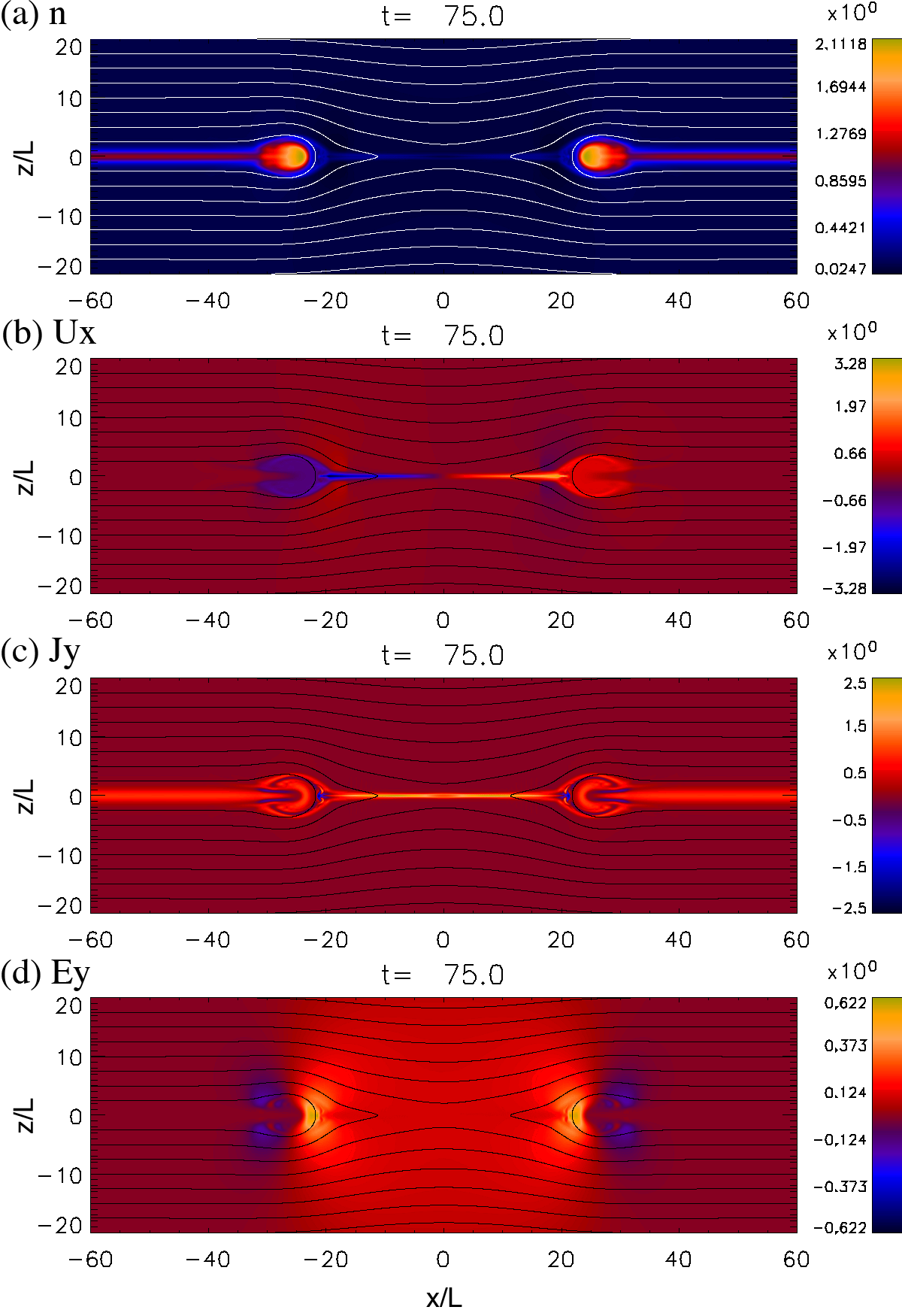}
\fi
\caption{
Snapshots of run L3 at $t/\tau_c=75$ in the $x$-$z$ two-dimensional plane.
(\textit{a}) The plasma proper density $n/n_0$,
(\textit{b}) the $x$-component of the 4-velocity of the plasma flow $u_x/c$,
(\textit{c}) the out-of-plane electric current $j_y/j_0$, and
(\textit{d}) reconnection electric field $E_y/B_0$.
The solid lines show magnetic field lines.
\label{fig:snap75}}
\end{center}
\end{figure}

\section{SIMULATION RESULTS}
\subsection{Evolution overview}

In this section,
we overview the system evolution of
our reference run (run L3) in detail.
Due to the trigger field,
magnetic reconnection occurs
around the center of the simulation domain.
Plasma outflows start to travel into the $\pm x$ directions
from the center, while inflows come from the $\pm z$ directions.
The panels in Figure \ref{fig:snap75} show
various physical properties
at $t/\tau_c=75$ in the normalized unit:
the plasma proper density $n$,
the plasma 4-velocity $u_x$,
the electric current $j_y=2 q_p n u_y$, and
the reconnection electric field $E_y$.
Since reconnection outflows eject a lot of plasmas,
we see dense plasma islands (plasmoids)
around $x/L \sim \pm25$-$30$ (Figure \ref{fig:snap75}\textit{a}).
The reconnection outflow jets become very fast, up to $u_x\sim3.28c$
(Figure \ref{fig:snap75}\textit{b}).  
The $u_x$ profile shows
a characteristic crab claw structure in the plasmoid region,
because
the dense current sheet plasmas exist
around the neutral plane ($z\sim 0$).
We also see weak reverse flows around $x/L \sim \pm 20$
after plasmoid passing.
In the reconnecting region, there is a thin central current layer,
and its peak current is $2.5$ times larger than the initial state
(Figure \ref{fig:snap75}\textit{c}).
Similar enhancement is often seen in classical nonrelativistic models.
There are rather complicated current structures
inside the plasmoids.
At this stage
the out-of-plane electric field (or the reconnection electric field)
is well developed (Figure \ref{fig:snap75}\textit{d}).
The typical amplitude is $E_y/B_0\sim 0.1$ over the reconnection region.
In addition, the electric field is enhanced $E_y/B_0\sim 0.6$
around $x/L\sim \pm20$-$25$,
where reconnected magnetic flux $B_z$ is accumulated.
The energy and momentum of these enhanced fields are converted to
those of the downstream plasmas.
We also note that this pileup region
plays an interesting role as particle accelerator \citep{claus04,zeni07}.
In general, the magnetic topology, electric field properties,
and spatial distribution of plasma properties
are sufficiently consistent with previous reconnection studies
by PIC or MHD simulations.
The system evolution is similar to the Sweet--Parker reconnection
which features a single current sheet,
although the reconnection grows fast.

\begin{figure}[htbp]
\begin{center}
\ifjournal
\includegraphics[width={0.85\columnwidth},clip]{f2.eps}
\else
\includegraphics[width={0.85\columnwidth},clip]{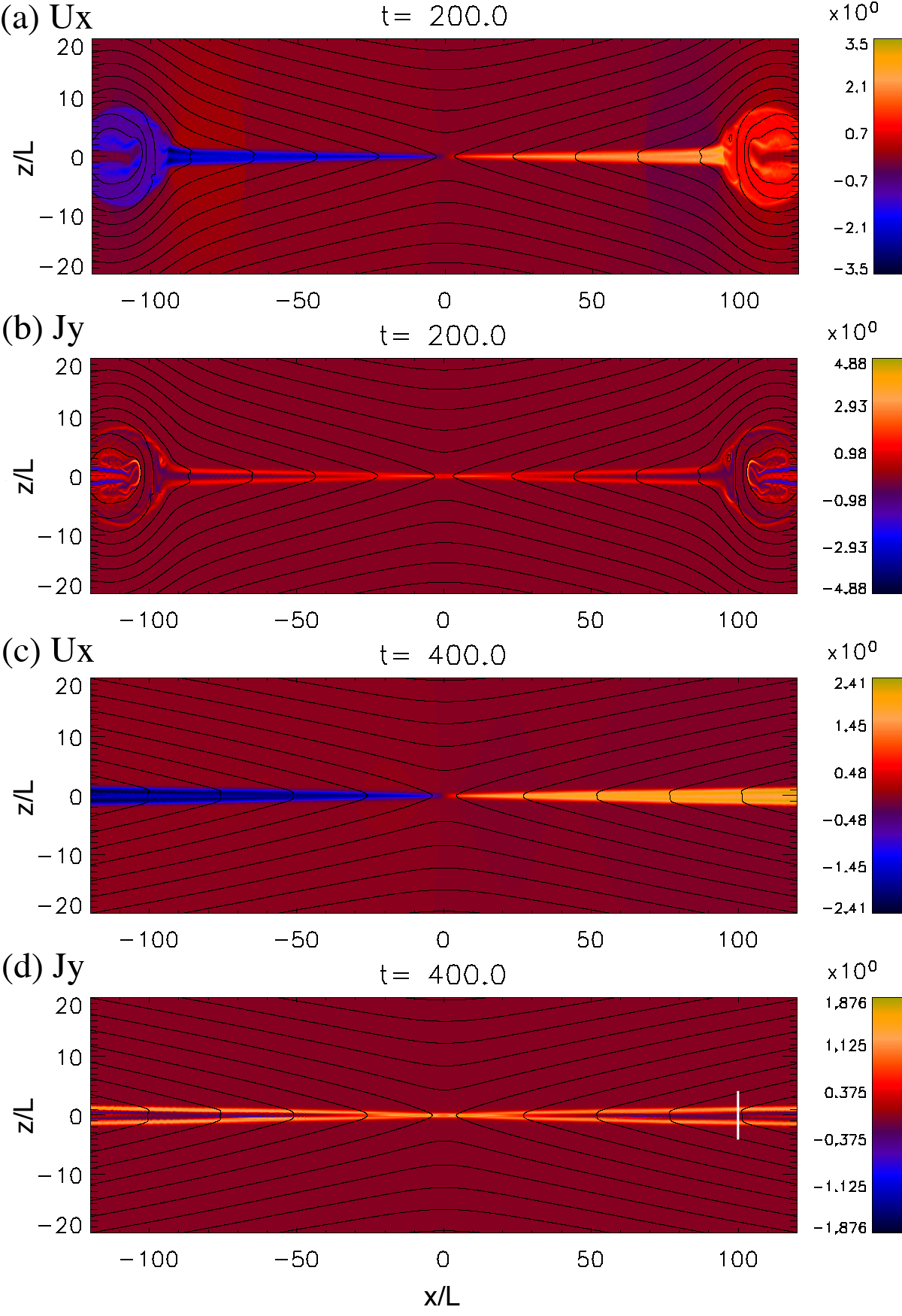}
\fi
\caption{
Large-scale snapshots of run L3 in the $x$-$z$ two-dimensional plane.
(\textit{a}) The $x$-component of the plasma 4-velocity $u_x/c$ at $t/\tau_c=200$,
(\textit{b}) the out-of-plane current $j_y/j_0$ at $t/\tau_c=200$,
(\textit{c}) the $x$-component of the plasma 4-velocity $u_x/c$ at $t/\tau_c=400$, and
(\textit{d}) the out-of-plane current $j_y/j_0$ at $t/\tau_c=400$.
The black lines show the magnetic field lines.
We later discuss the properties along the white line ($x/\L=100$)
in panel (\textit{d}) in Section \ref{sec:slowshock}.
\label{fig:snap200}}
\end{center}
\end{figure}

After the initial phase, the reconnection continues and
plasmoids travel into the $\pm x$-directions.
The top two panels in Figure \ref{fig:snap200} show
late-time snapshots at $t/\tau_c=200$.
At this stage, plasmoids start to reach
the outflow boundaries,
as we see in the $u_x$ profile (Fig.~\ref{fig:snap200}\textit{a}).
Note that the entire domain is presented in the $x$ direction.
The fastest flows $u_x\sim 3.5c$ are found
at $x/L\sim\pm 90$ along the outflow line,
where the outflow channels are connected to the plasmoids.
An important feature is found in the electric current profile
(Figure \ref{fig:snap200}\textit{b}).
From the central $X$-type region to the downstream region,
the current layers are now bifurcated.
The bifurcation starts around $x/L=\pm 40$ at $t/\tau_c=100$-$125$. 
We think these current layers are
a signature of the Petschek-type steady reconnection,
which enables faster energy conversion,
and we analyze their structure in a later section (see \ref{sec:slowshock}).
Interestingly, we see weak ``reverse currents''
between the two current layers.
The current structures inside the plasmoids
become further complicated,
including the interaction with boundaries.

Since we employ the open boundary condition,
plasmoids and reconnection outflows pass through the $x$-boundaries.
Since plasmas and magnetic field lines are continuously supplied
from the inflow open boundaries at $z=\pm L_z$,
the reconnection still continues, and therefore
the system evolves further.
Importantly, the system grows into
a steady state reconnection structure
after the plasmoids have left.
The bottom two panels in Figure \ref{fig:snap200} show
the snapshots of a very late stage at $t/\tau_c=400$.
Now the outflow channels (Figure \ref{fig:snap200}\textit{c})
between Petschek-type current layers (Figure \ref{fig:snap200}\textit{d})
are found all over the $x$ direction.
The distance between the two current layers is $\sim2.5$-$3L$
at the outflow boundary ($x/L=L_x=120$).
Thus, the slope angle of the current layer is very small,
compared to a typical slow-shock angle of $\theta \sim 0.1$
in nonrelativistic Petschek reconnection.
The magnetic field line structure is very smooth
over the entire simulation domain.
We find that these current layers remain stable
for a relatively long time.


\begin{figure}[htbp]
\begin{center}
\ifjournal
\includegraphics[width={0.75\columnwidth},clip]{f3a.eps}
\includegraphics[width={0.75\columnwidth},clip]{f3b.eps}
\includegraphics[width={0.75\columnwidth},clip]{f3c.eps}
\includegraphics[width={0.75\columnwidth},clip]{f3d.eps}
\includegraphics[width={0.75\columnwidth},clip]{f3e.eps}
\else
\includegraphics[width={0.75\columnwidth},clip]{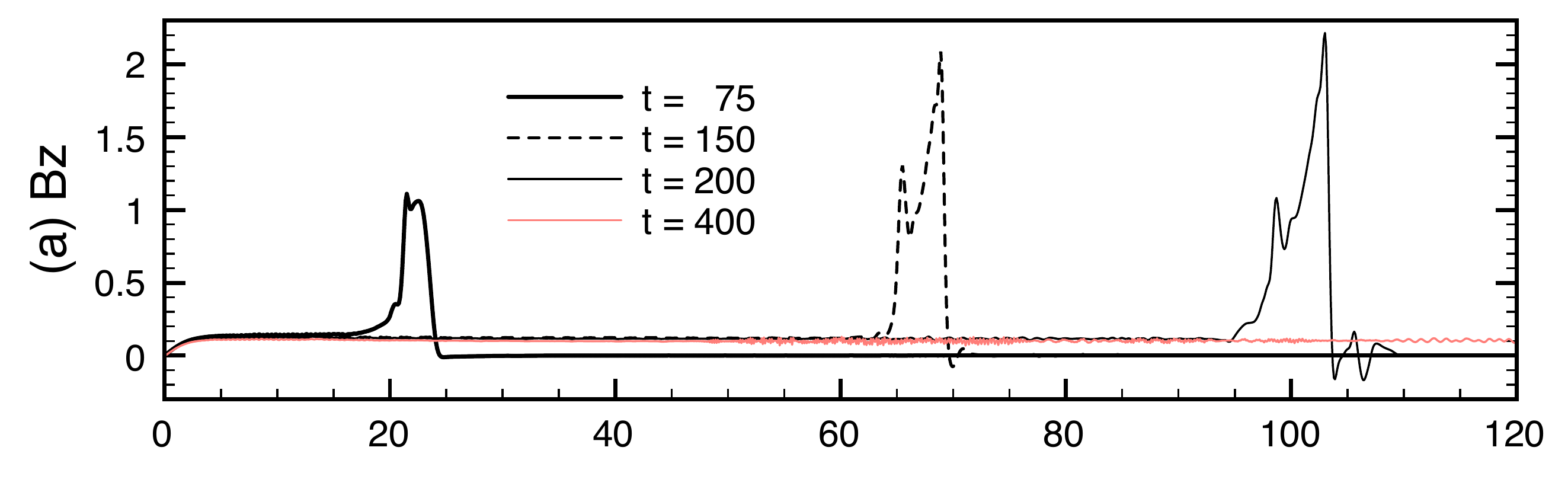}
\includegraphics[width={0.75\columnwidth},clip]{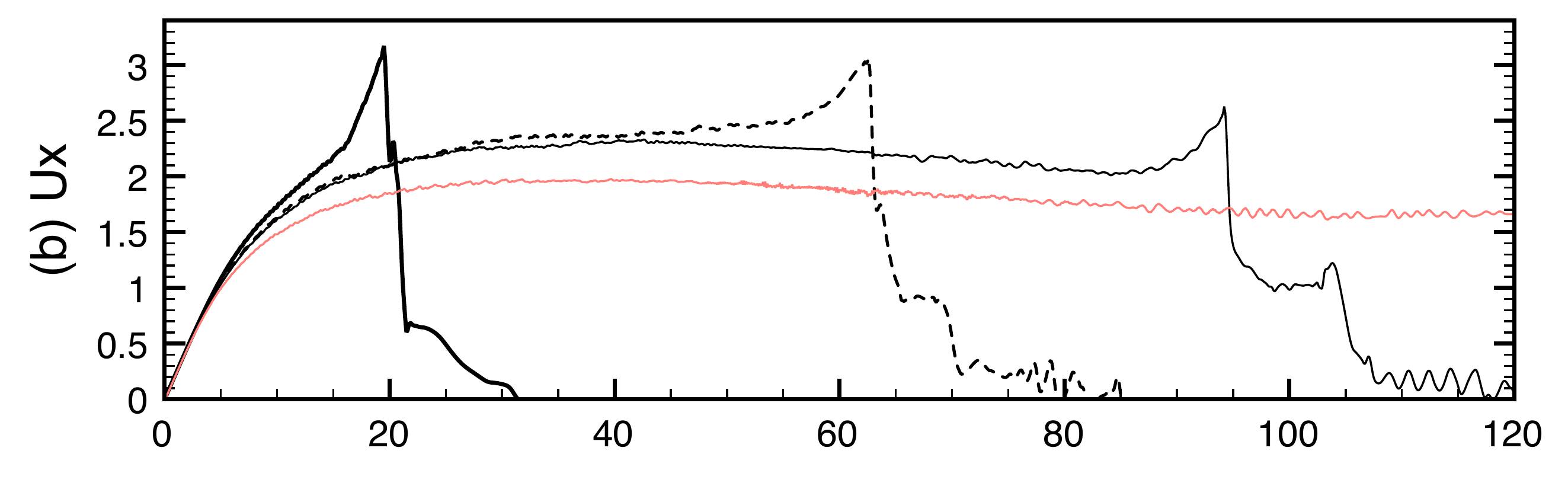}
\includegraphics[width={0.75\columnwidth},clip]{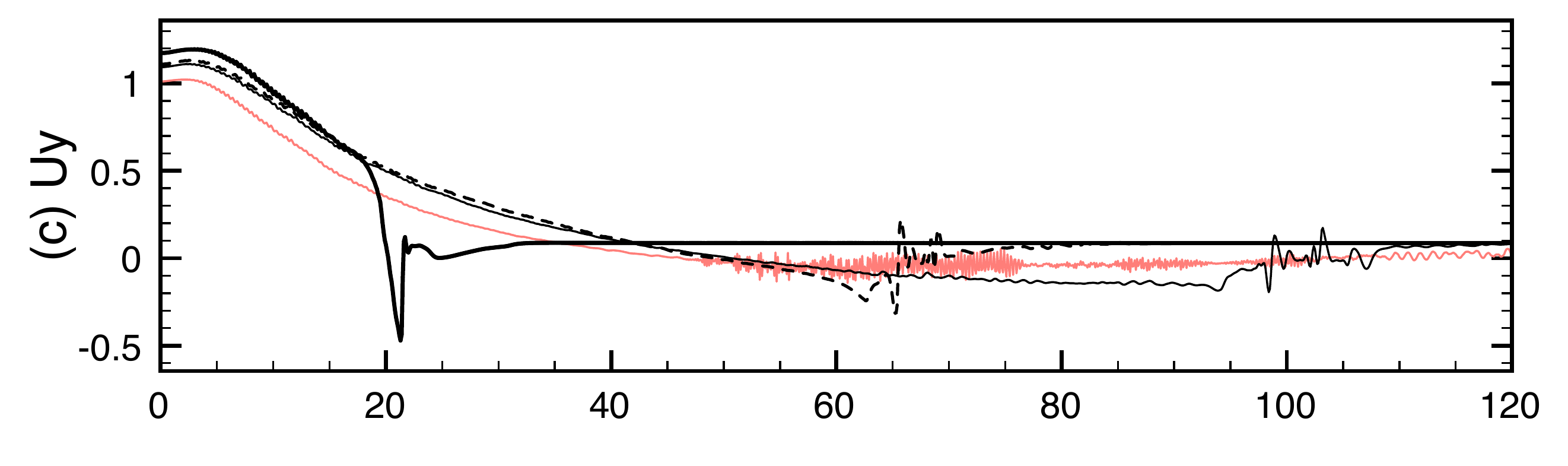}
\includegraphics[width={0.75\columnwidth},clip]{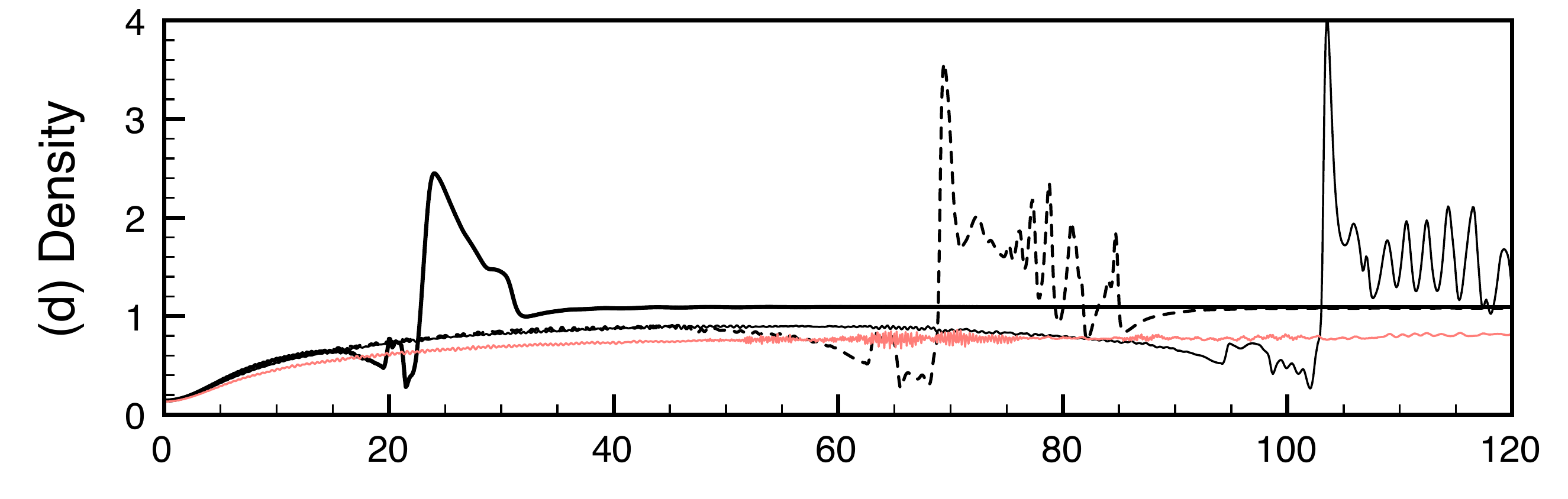}
\includegraphics[width={0.75\columnwidth},clip]{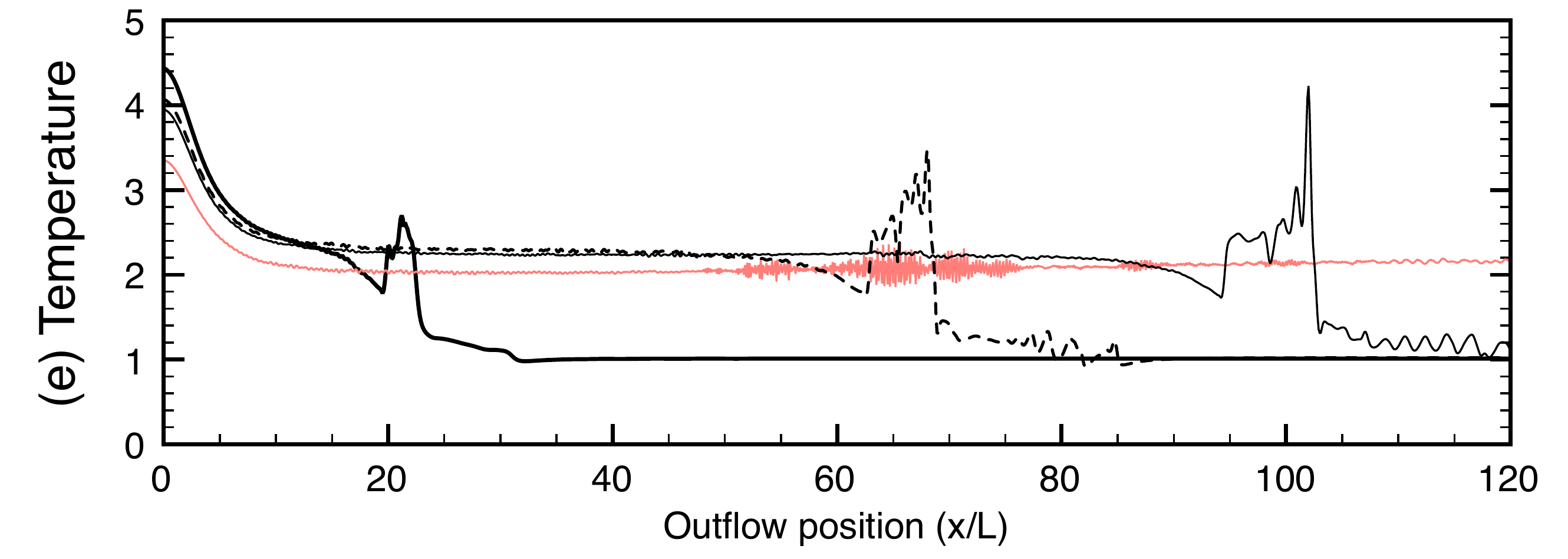}
\fi
\caption{
Temporal evolution of physical properties
along the outflow line ($z=0$) in run L3.
(\textit{a}) The vertical magnetic field $B_z/B_0$,
(\textit{b}) the outflow component of the positron 4-velocity $u_x/c$,
(\textit{c}) the out-of-plane component of the positron 4-velocity $u_y/c$,
(\textit{d}) the normalized plasma number density $\gamma n/n_0$, and
(\textit{e}) the normalized plasma temperature $p/(nmc^2)$.
\label{fig:outflow}}
\end{center}
\end{figure}

Next, we investigate the structures of the outflow region
in more detail.
Figure \ref{fig:outflow} presents the temporal evolution of
physical properties along the outflow line ($z=0$)
in the normalized units.
We compare the three nonsteady stages ($t/\tau_c=75,150,200$) in black lines,
and the late-time steady stage ($t/\tau_c=400$) in red.
The vertical magnetic field $B_z$
(Figure \ref{fig:outflow}\textit{a}) is
a reconnected component of magnetic field lines.
It is zero at the $X$-point, and
it remains at constant level of $\sim0.1B_0$ inside the outflow channel.
Strong peaks are the pileup regions,
where the reconnected field lines are piled up
in front of the dense plasmas.
As discussed, the electric fields are also enhanced there.
Such a powerful magnetic pileup and the relevant motional electric fields
are signatures of fast magnetic reconnection.
The pileup is so strong that
several discontinuities appear near the pileup regions.
For example, at the upstream side of the pileup region,
the outflow speed $u_x$ becomes very fast but it suddenly drops
(Figure \ref{fig:outflow}\textit{b}).
On the other hand, there is a strong jump in $B_z$ at the downstream side of the pileup region,
although the velocity jump is not so clear.
We think they are
the tangential discontinuity or a weak shock (the downstream one)
and the relevant reverse fast shock (the upstream one).
In the later stages ($t/\tau_c \gtrsim 150$),
the system starts to suffer from numerical noises
in the downstream side of the plasmoids,
as seen in the velocity profile or
in the density profile (Figure \ref{fig:outflow}\textit{d}).
These noises go away as plasmoids pass through the outflow boundaries.
Importantly, we find that
the out-of-plane 4-velocity $u_y$ is not negligible
over the relatively large region $|x/L| \lesssim 20$
(Figure \ref{fig:outflow}\textit{c}).
Since $u_y$ is coupled with in-plane components $u_x$ and $u_z$,
this immediately implies that
the conventional one-fluid MHD approximation breaks down
and that the two-fluid approximation is essential there.
At $t/\tau_c=200$, $u_y$ becomes negative around $x/L \sim 80$-$90$.
This stands for the negative current between the Petschek-type current layers.
The bottom panel (Figure \ref{fig:outflow}\textit{e}) shows
the plasma temperature $T=p/nmc^2$.
It is very large at the reconnecting $X$-point,
and also $(T\sim 2nmc^2)$ inside the outflow channel.
The typical Lorentz factors in the outflow region are
$\gamma \sim 2.4 \pm 0.2$ ($t/\tau_c=200$) and $2.1 \pm 0.1$ ($t/\tau_c=400$),
They are comparable with an Alfv\'{e}nic value $(1+\sigma_{\varepsilon,in})^{1/2}=2.2$.

\begin{figure}[htbp]
\begin{center}
\ifjournal
\includegraphics[width={0.93\columnwidth},clip]{f4.eps}
\else
\includegraphics[width={0.93\columnwidth},clip]{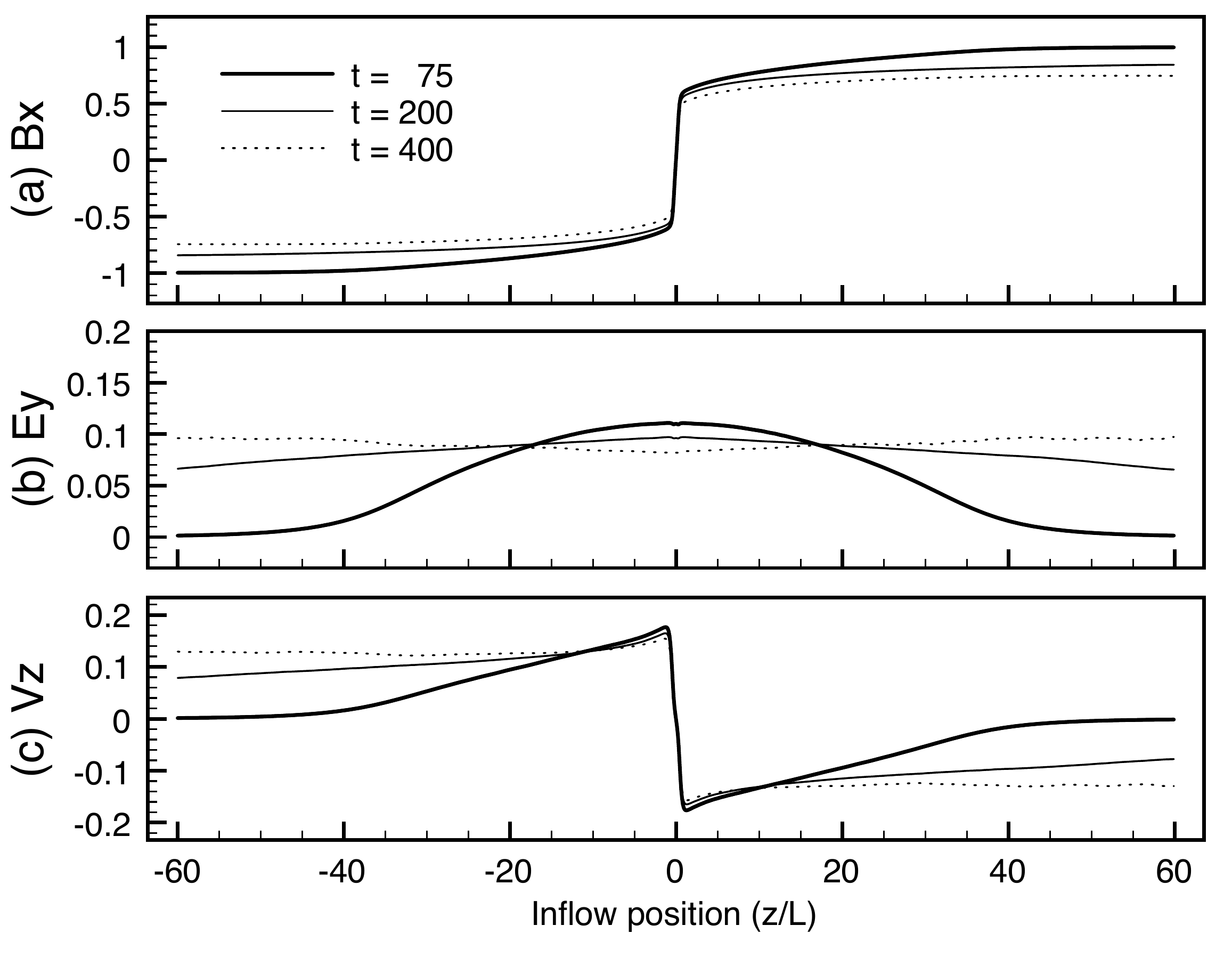}
\fi
\caption{
Temporal evolution of physical properties
along the inflow line ($x=0$) in run L3.
(\textit{a}) The antiparallel magnetic field $B_x/B_0$,
(\textit{b}) the reconnection electric field $E_y/B_0$, and
(\textit{c}) the inflow positron velocity $v_z/c$.
\label{fig:inflow}}
\end{center}
\end{figure}

It is important that
the late-time profiles at $t/\tau_c=400$
(indicated by the red lines in Figure \ref{fig:outflow})
are quite similar to the earlier profiles.
This tells us that
the late-time structure (Figures \ref{fig:snap200}\textit{c} and \ref{fig:snap200}\textit{d})
is a very good prediction of the steady state profile.
We still see a numerical noise around $x/L \sim 60$-$70$.
This is because
this outflow channel is located
in the downstream side of the shock-type region.
As discussed, the outflow channel is located at
the downstream side of the two current layers.

Next, we visit the physical properties along the inflow line ($x=0$).
From Figure \ref{fig:inflow}\textit{a} we know that
the reconnection starts to consume the antiparallel magnetic field $B_x$,
but it goes down to an asymptotic level of
$B_x \sim \pm 0.74B_0$ due to the open boundary condition.
The field reversal is localized in the narrow region around $z\sim 0$. 
The reconnection electric field $E_y$ grows as the system evolves
(Figure \ref{fig:inflow}\textit{b}).
At the later stages, it becomes constant over the simulation domain.
This tells us that our open-boundary condition works excellently.
Also, plasma inflow remains at the constant level around the center
(Figure \ref{fig:inflow}\textit{c}).
This us tells that the reconnection constantly goes on,
consuming outside plasmas and magnetic fields
at the constant rate.

\begin{figure}[htbp]
\begin{center}
\ifjournal
\includegraphics[width={\columnwidth},clip]{f5.eps}
\else
\includegraphics[width={\columnwidth},clip]{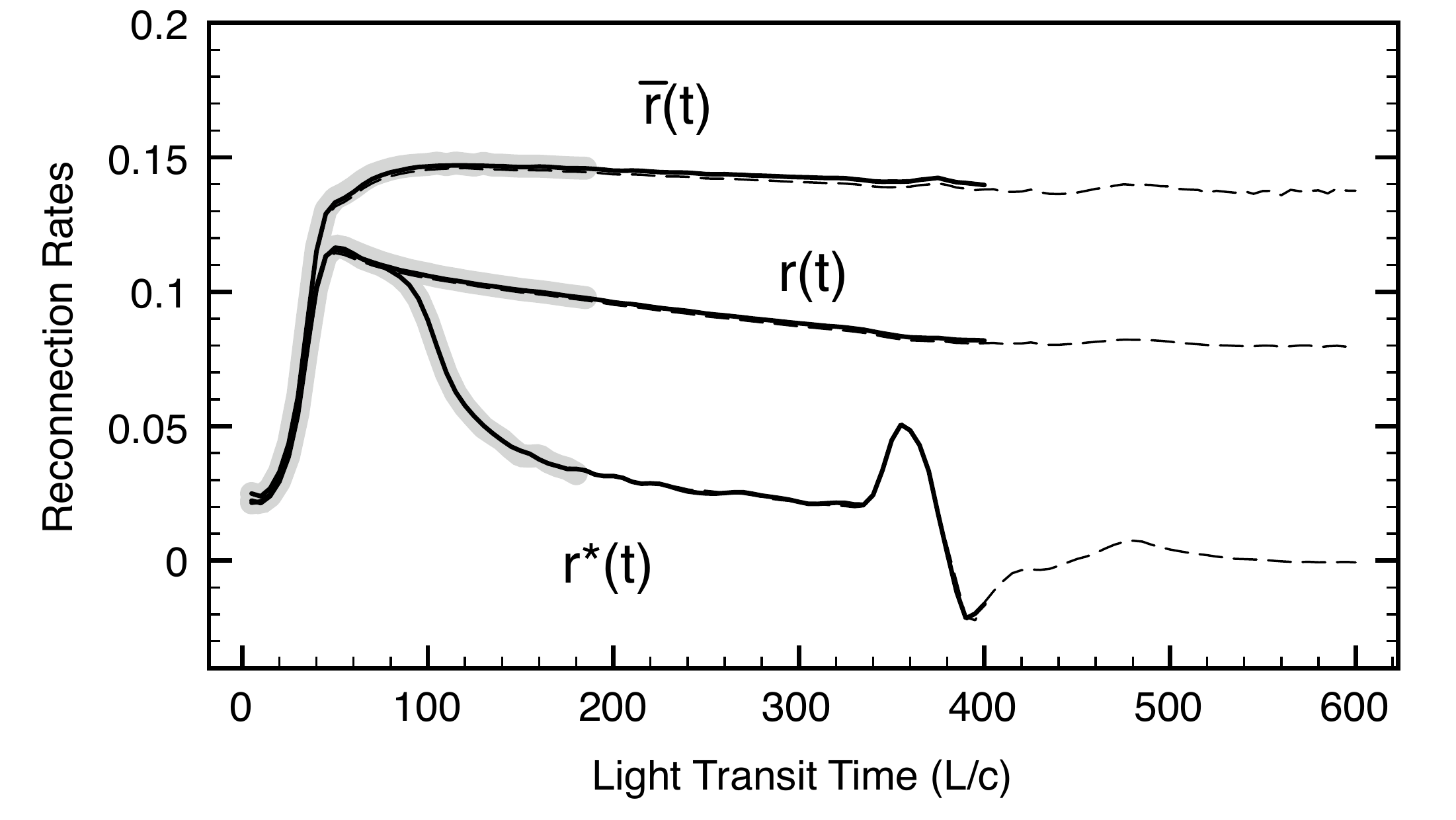}
\fi
\caption{
Time evolution of the reconnection rates
in runs
L3 (\textit{solid lines}),
M3 (\textit{dashed lines}),
and XL3 (\textit{gray thick lines}).
The raw reconnection rate $r(t)=E_y/B_0$ at the $X$-point,
the normalized reconnection rate $\bar{r}(t)=cE_y/[c_{A,in'}B_{x,in'}]$,
and the flux consumption rate $r^*(t)$ are presented.
\label{fig:longterm}}
\end{center}
\end{figure}

The amplitude of the reconnection electric field
\begin{eqnarray}
r(t)=E_y / B_0
\end{eqnarray}
at the $X$-point
is one of the most important parameters in a magnetic reconnection.
This measures how fast the system transports the magnetic flux into the $X$-point,
or how fast the reconnection consumes the upstream magnetic energy.
It is often referred as the ``reconnection rate''
in various normalized form.
Following convention,
we used the following reconnection rate,
because reconnection outflow speed is often
approximated by the upstream Alfv\'{e}n speed:
\begin{eqnarray}
\bar{r}(t)=\frac{c E_y }{ c_{A,in'} |B_{x,in'}| }.
\end{eqnarray}
Here the subscript $in'$ denotes the inflow properties measured at $z/L=20$.
The time evolution of $r(t)$ and $\bar{r}(t)$ is
presented in Figure \ref{fig:longterm}.
In addition to the reference run L3,
two other runs M3 and XL3 (similar runs with difference resolutions) are
overplotted in order to check the convergence of the simulation:
three are in excellent agreement. 
The normalized rate $\bar{r}(t)$ is larger than the raw rate $r(t)$,
mainly because the inflow magnetic field
$B_{x,in'}$ decreases over time (Figure \ref{fig:inflow}\textit{a}).
We see that both the rates remain stable throughout the system evolution.
Indeed, the normalized rate remains constant: $\bar{r}(t)\sim 0.14$.

The other quantity $r^*(t)$ is
the time derivative of
the accumulated magnetic flux along the inflow line
\begin{eqnarray}
r^*(t)=-\frac{d}{dt}\int_0^{L_z} B_x ~dz
\end{eqnarray}
Because of the discrete sampling time,
the calculated value is rather crude, but
is useful enough to validate the simulation results.
In the early stage,
both $r(t)$ and $r^*(t)$ are in excellent agreement.
They do not agree after $t/\tau_c > 80$,
because the magnetic flux enters from the open inflow boundaries.
During $ t/\tau_c \sim 340$-$400$,
$r^*(t)$ exhibits strange behavior.
We confirmed that this is a boundary effect.
Since plasmoid passes through the outflow boundaries
around $ t/\tau_c \sim 200$-$250$,
perturbation travels from there
as a light wave or a fast Alfv\'{e}n wave.
The waves from the two outflow boundaries
arrived at the center of the inflow boundaries.
Since two waves carry outward energy flux,
the incoming magnetic flux temporally slows down, but
the system adjusts itself and
it goes back to the quasi-steady state after $t/\tau_c > 400$.
Note that the final asymptotic value $r^*(t) = 0$ indicates
the steady evolution.

\subsection{Case studies\label{sec:cases}}

In this section, we compare various simulation runs,
focusing on the composition of the typical upstream energy flow $\sigma_{\varepsilon,in}$.
As presented in Table \ref{table},
this parameter is mainly controlled by the upstream plasma density $n_{in}/n_0$.
The magnetically dominated cases of $\sigma_{\varepsilon,in} \gg 1$
(``high-$\sigma$'' runs) are of strong astrophysical interest,
while
plasma-dominated cases of $\sigma_{\varepsilon,in} < 1$ (low-$\sigma$ runs)
can be compared with nonrelativistic reconnection studies.

\begin{figure}[htbp]
\begin{center}
\ifjournal
\includegraphics[width={\columnwidth},clip]{f6.eps}
\else
\includegraphics[width={\columnwidth},clip]{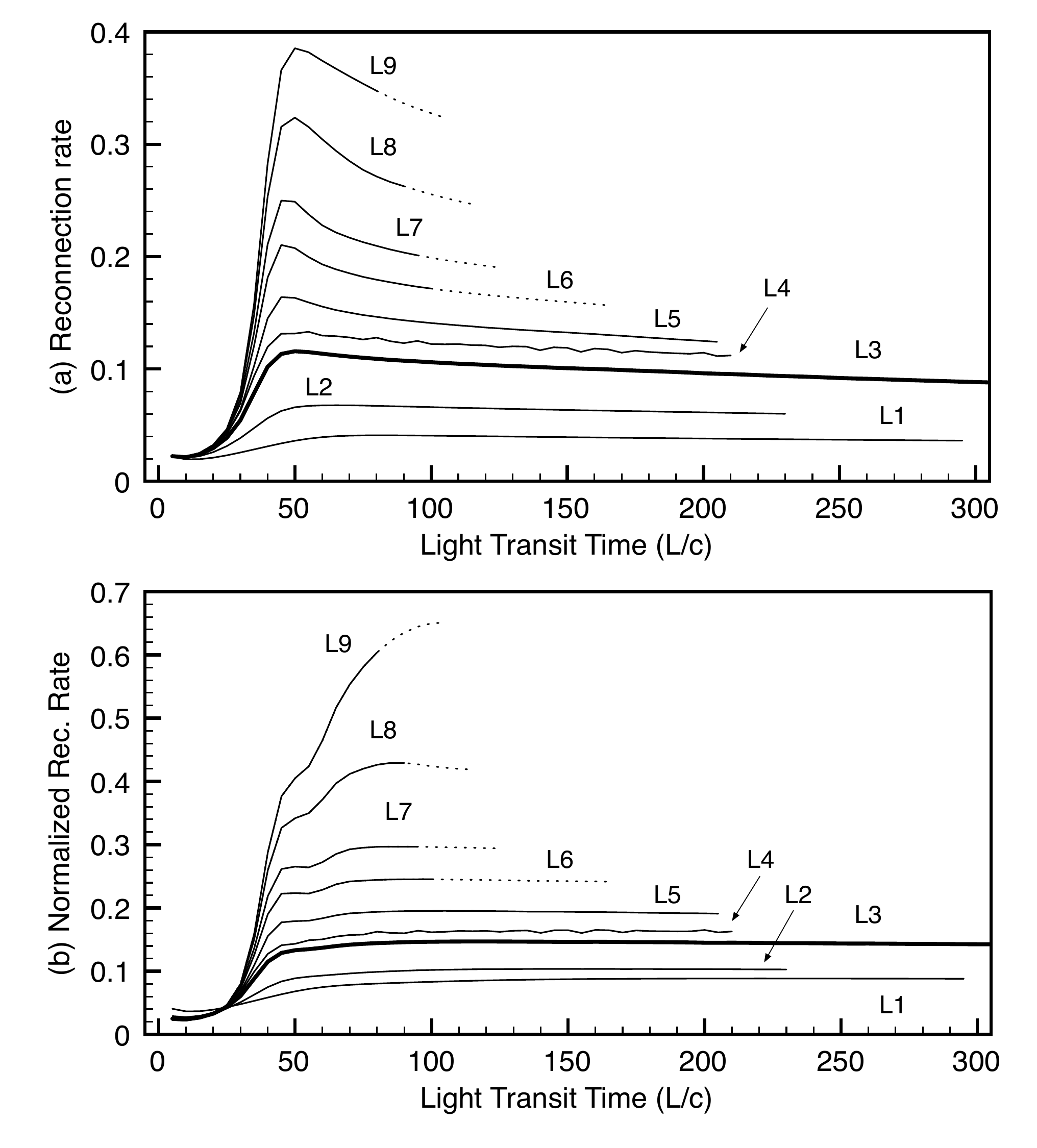}
\fi
\caption{Temporal evolution of the reconnection electric field
(reconnection rate) at the $X$-point;
(\textit{a}) 
the raw reconnection rate $r(t)=E_y/B_0$ and
(\textit{b})
the normalized reconnection rate $\bar{r}(t)=cE_y/[c_{A,in'}B_{x,in'}]$.
The reference run L3 is presented in thick lines.
The dotted lines contain negative mass density.
\label{fig:recrate}}
\end{center}
\end{figure}

Figure \ref{fig:recrate} presents
the reconnection rate $r(t)$ and
the normalized reconnection rate $\bar{r}(t)$
for simulation runs L1-L9 in Table \ref{table}. 
Generally speaking, the lower-$\sigma$ runs L1-L5 last relatively long time.
Their time duration $t/\tau_c \sim 200$-$300$ is related to
the plasmoid collisions.
Therefore, the system has enough time to evolve to the Petschek-type reconnection,
and we recognize Petschek-type current layers in these runs.
We will visit the physical property of typical low-$\sigma$ run (run L1)
later in this section.
Run L4 is the cold inflow counterpart of run L3;
it uses the same parameters as run L3,
except for the upstream plasma pressure.

The higher-$\sigma$ runs L6-L9 become unstable,
and they stop before $t/\tau_c \lesssim 100$.
The numerical problem occurs around the plasmoids
in the reconnection outflow front.
As discussed in Section 3.1, there are discontinuities
both in the upstream and the downstream of
the magnetic pileup region.
Since our numerical scheme (Lax--Wendroff scheme) is not ideal for shocks,
we suffer from numerical noise at these discontinuities.
Since the magnetic energy dominates the plasma energy in these runs,
even small noises in the electromagnetic fields become crucial to fluid properties,
and then the physically valid solution often collapses.
In the dotted line region, we continue simulations
even though we observe small negative mass in the edge of the plasma outflow,
until our equation solver fails to find the mathematical solution.
Since the numerical error occurs near the plasmoid,
we think they show the right evolution for a while ($20 \sim 30\tau_c$),
until the unphysical information comes back to the $X$-point.
We find that the normalized rate $\bar{r}(t)$ (Figure \ref{fig:recrate}\textit{b})
is a better measure of the reconnection evolution,
because it looks reasonably flat in higher-$\sigma$ runs.
However, we will only consider the times prior to
the occurrence of negative density.

\begin{figure}[htbp]
\begin{center}
\ifjournal
\includegraphics[width={\columnwidth},clip]{f7.eps}
\else
\includegraphics[width={\columnwidth},clip]{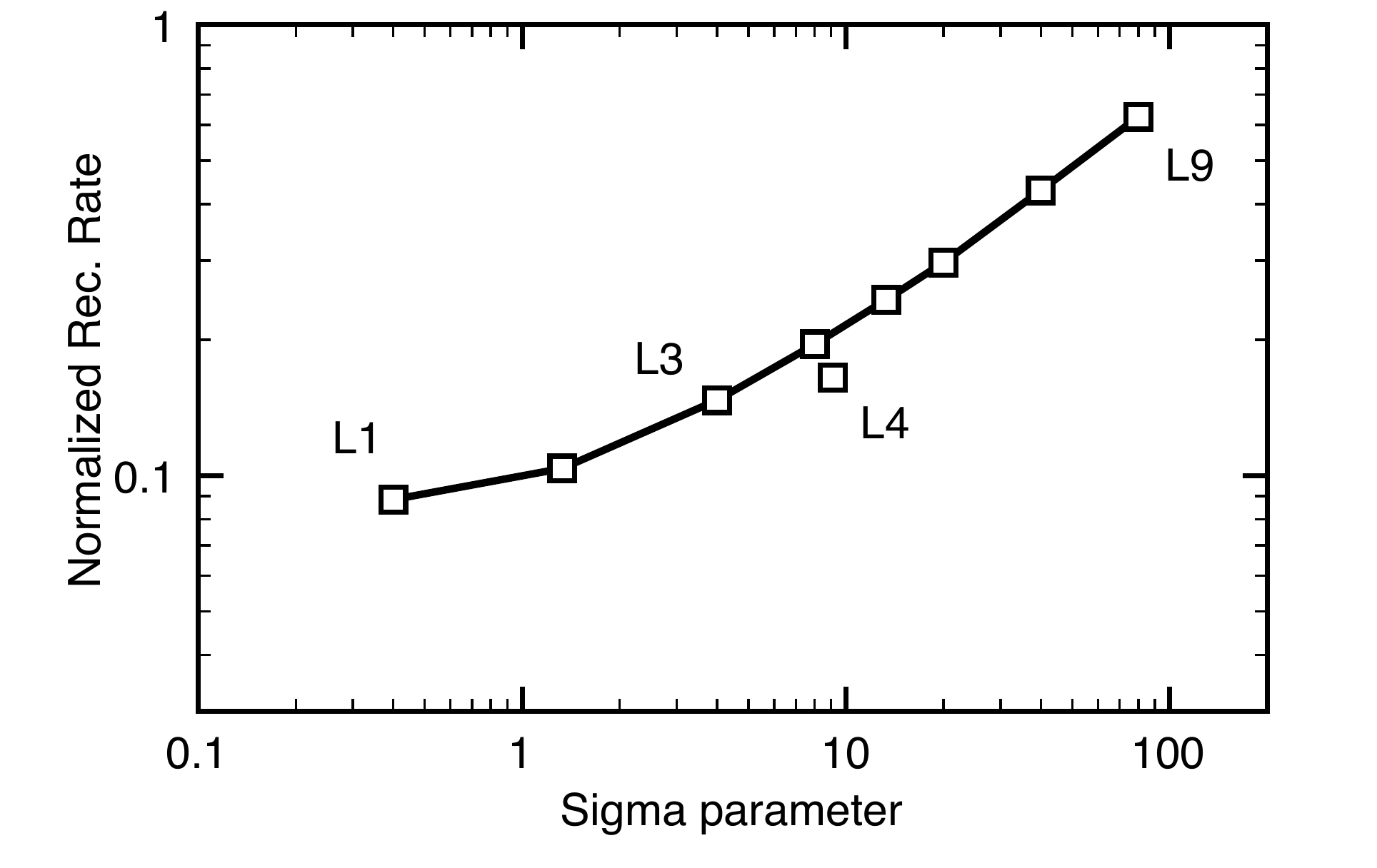}
\fi
\caption{
Dependence of the maximum reconnection rate $\bar{r}(t)$,
as a function of the initial upstream parameter $\sigma_{\varepsilon,in}$.
\label{fig:maxrate}}
\end{center}
\end{figure}

As a general trend, we find that
the reconnection rate becomes higher
as the inflow density goes down, or
as the parameter $\sigma_{\varepsilon,in}$ increases.
Figure \ref{fig:maxrate} also shows
the maximum reconnection rate $\bar{r}(t)$ in the simulation runs,
as a function of the initial upstream $\sigma_{\varepsilon,in}$ parameter.
In the limit of $\sigma_{\varepsilon,in}<1$,
the reconnection rate is asymptotic to $\sim$0.1.
This is consistent with
many studies on the nonrelativistic Petschek reconnection,
whose the reconnection rate is known to be $\sim$0.1.
On the other hand, the rate constantly increases
as the parameter $\sigma_{\varepsilon,in}$ increases.
It is striking that the reconnection rate becomes closer to $\sim$1,
because the rate of one is the upper limit of magnetic dissipation.

\begin{figure}[htbp]
\begin{center}
\ifjournal
\includegraphics[width={0.85\columnwidth},clip]{f8.eps}
\else
\includegraphics[width={0.85\columnwidth},clip]{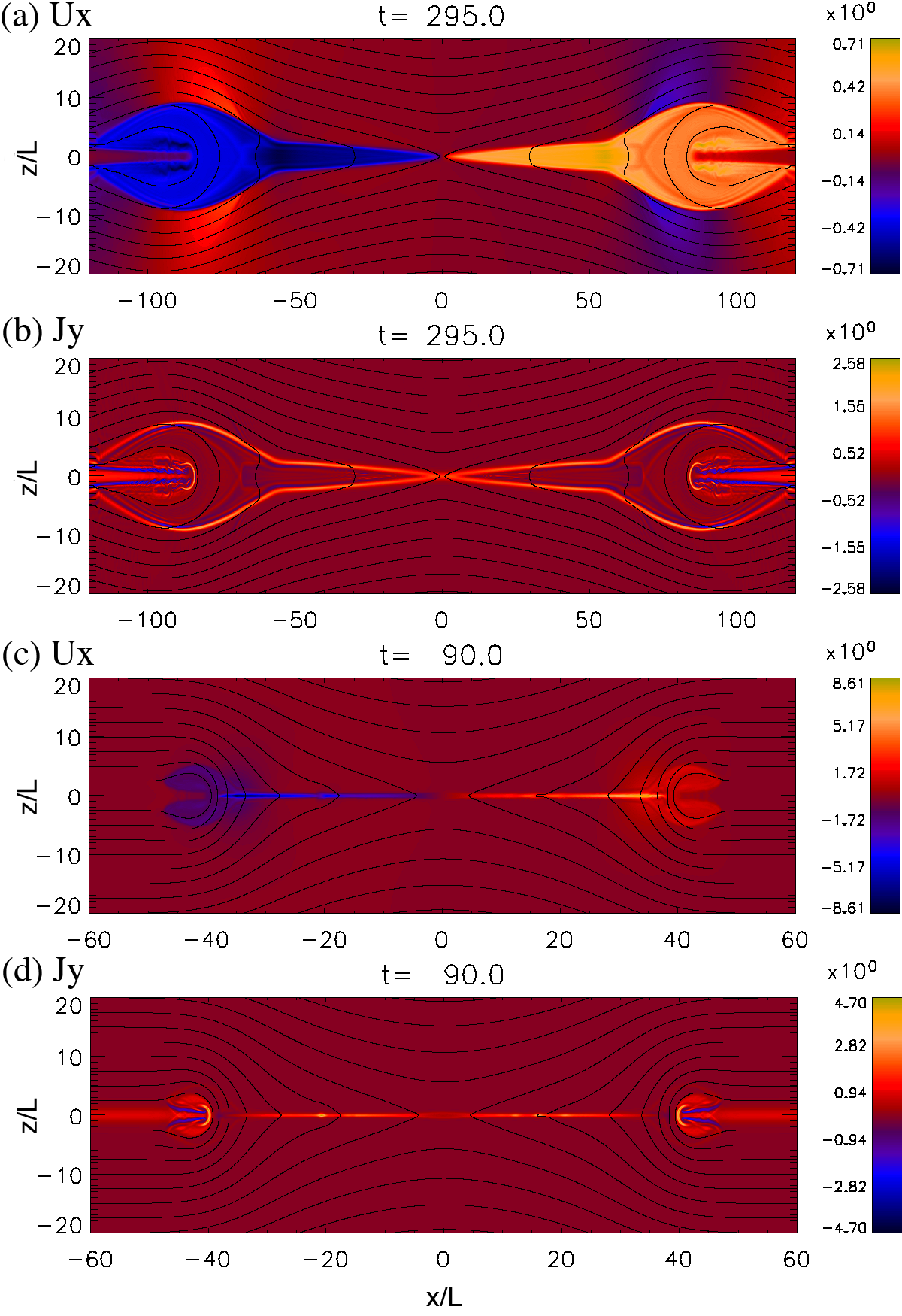}
\fi
\caption{
Top two panels: large-scale snapshots of run L1 at $t/\tau_c=295$.
(\textit{a}) The $x$-component of the plasma 4-velocity $u_x/c$, and
(\textit{b}) the out-of-plane current $j_y/j_0$.
Bottom two panels: snapshots of run L8 at $t/\tau_c=90$.
(\textit{c}) The $x$-component of the inflow plasma 4-velocity $u_x/c$, and
(\textit{d}) the out-of-plane current $j_y/j_0$.
\label{fig:cases}}
\end{center}
\end{figure}

We briefly visit the global properties of low-$\sigma$ runs.
Top two panels in Figure \ref{fig:cases} presents
the late time snapshots at $t/\tau_c=295$ in run L1.
Compared with the other higher-$\sigma$ runs,
the system evolution is rather slower
due to the slow reconnection outflow.
The typical outflow speed $\sim$0.5c ($0.57c$ at maximum) is
consistent with the original upstream Alfv\'{e}n speed of $0.594c$.
In the current profile (Figure \ref{fig:cases}\textit{b}),
we find Petschek-type current layers
and the angle between current layers look wider than the reference run L3.
Another current layer surrounding the plasmoid is very clear, too.
These signatures are well observed in plasmoid
in nonrelativistic ion-electron plasmas.
Unfortunately, we do not obtain
long-term steady profile after the boundary collision, 
because the run stops immediately after this stage.

The bottom two panels in Figure \ref{fig:cases} show
snapshots of the second most extreme case, run L8.
In the outflow profile (Figure \ref{fig:cases}\textit{c}),
we see that the outflow channel is narrower than the slower counterparts.
At the edge of the Sweet--Parker outflow jets,
the outflow 4-velocity becomes further relativistic, $u_x/c \sim \pm 8.6$,
and the maximum Lorentz factor in the system is up to $\sim$9.
The current structure remains in a single current
(Figure \ref{fig:cases}\textit{d})
at least at this stage.
In the very thin current layer,
there are small seeds of secondary tearing islands
(e.g. a bright spot at $x/L\sim-20$ in the current profile;
Figure \ref{fig:cases}\textit{d}).

\begin{figure}[htbp]
\begin{center}
\ifjournal
\includegraphics[width={\columnwidth},clip]{f9.eps}
\else
\includegraphics[width={\columnwidth},clip]{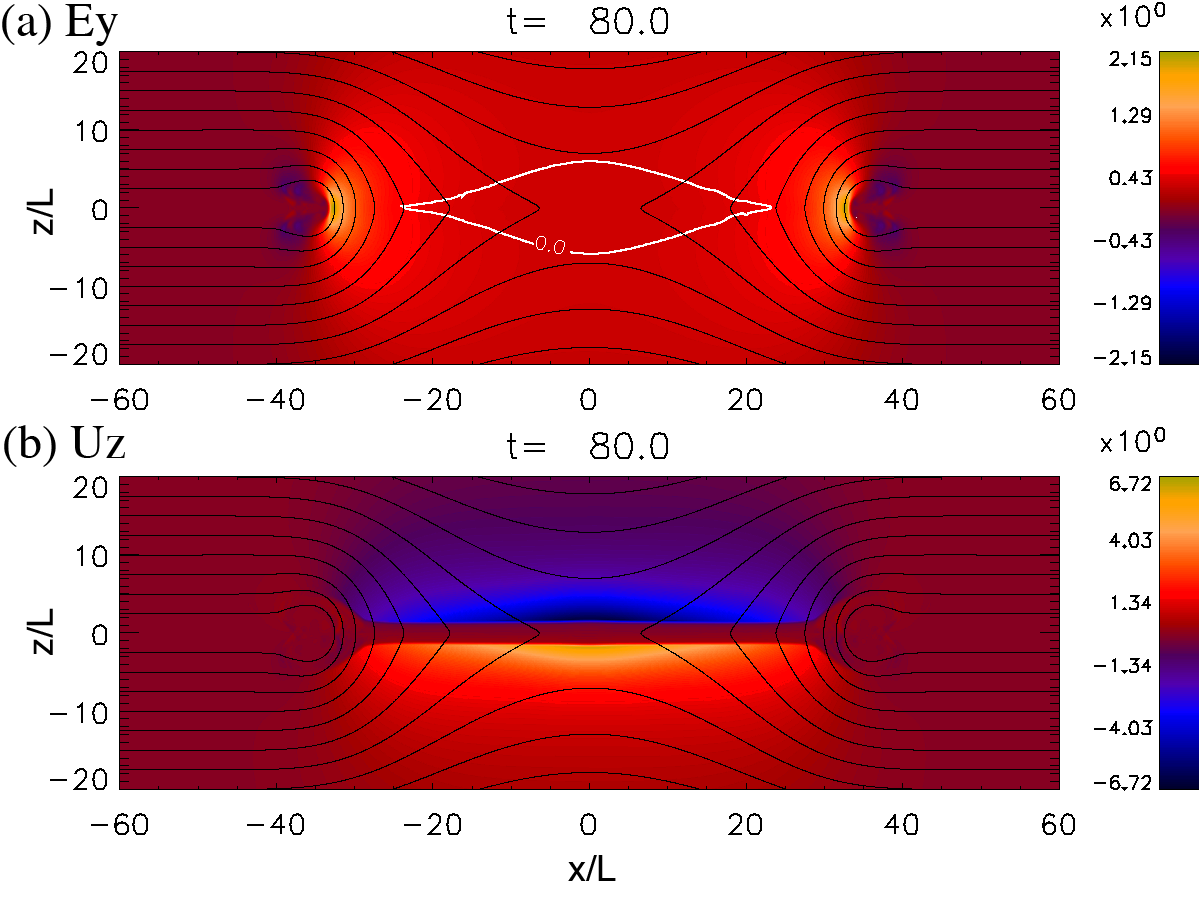}
\fi
\caption{
Snapshots of run L9 at $t/\tau_c=80$.
(\textit{a}) The reconnection electric field $E_y/B_0$.
The white contour line indicates the region
where the Lorentz invariant $(E^2-B^2)$ is positive.
(\textit{b}) The $z$-component of inflow 4-velocity $u_z/c$.
\label{fig:L9}}
\end{center}
\end{figure}

Furthermore, the most extreme case (run L9) shows an interesting evolution.
Panels in Figure \ref{fig:L9} show characteristic properties at $t/\tau_c=80$,
just before we meet an unphysical solution at $t/\tau_c=80.8$.
Figure \ref{fig:L9}\textit{a} shows the reconnection electric field.
Importantly, its amplitude is $E_y/B_0\sim0.35$-$0.5$.
We indicate the ``electric-dominant'' region
where the Lorentz invariant $(E^2-B^2)$ is positive with the white line.
We generally observe such an electric-dominant region
at the closer vicinity of the $X$-point,
because the reconnection electric field $E_y$ remains finite,
while the magnetic field $|B|$ becomes zero at the $X$-point.
However, such an electric-dominant region is usually confined
in a very narrow region of the center of the reconnecting current sheet.
For example, in run L8, such a region is very thin around the neutral plane,
$-0.5 < z/L < 0.5$.
However, in run L9,
the electric field $E_y$ becomes so strong that
it even dominates the magnetic field
in a relatively large spatial region of $-24 < x/L < 24, -6 <z/L < 6$.
In response to a strong electric field,
we also find a super fast reconnection inflow (Figure \ref{fig:L9}\textit{b}).
The maximum momentum is up to $|u_z|/c\sim 6.7$, and
the maximum inflow velocity is up to $|v_z|/c\sim 0.936$.
Also, as seen in Figure \ref{fig:L9}\textit{b},
the reconnecting current layer becomes thicker $1$-$2L$,
while in other cases the central current layer always becomes thin $\lesssim 0.5L$.
Since the plasma temperature becomes hot $5$-$10 mc^2$ along the current sheet,
the kinetic scale increases by a factor of $5$-$10$ and then
it is comparable to the initial sheet thickness $L$.
Therefore, we may have to consider kinetic effects beyond the fluid approximation.
Indeed, an effective resistivity based on the kinetic effects
is a long-standing problem in reconnection physics (e.g., \citet{hesse99}).

Regarding the energy conversion rates,
we noticed that the magnetic pileup regions are also important in the relativistic runs.
As $\sigma$ increases, the pileup fields become more strong, and then
more energy is delivered to the downstream Harris sheet plasmas there.
On the other hand, in the current sheets and in the Petschek-type current layers,
the energy conversion rate seems to be proportional to the reconnection rate. 
However, unfortunately, we do not have sufficient simulation results to discuss
energy conversion in the high-$\sigma$ regime, which is of strong astrophysical interest.

\subsection{Petschek-type current layer\label{sec:slowshock}}

One of the most characteristic features of
the late-time evolution of reconnection
is the bifurcated Petschek-type current layers.
We observe such current layers in runs L1-L5.
In this section, we study
how these current layers are influenced by
the upstream energy composition $\sigma_{\varepsilon,in}$.
In the relativistic Petschek reconnection,
\citet{lyu05} examined the RMHD jump conditions across the slow shocks,
and he found that the slow-shock angle becomes narrow
when $\sigma_{m,in} \gg 1$. 
We examine our simulation results based on a similar theory. 
Since \citet{lyu05}'s original work
employs single-fluid RMHD model and
it neglects the inflow plasma pressure,
first, we construct complete jump conditions
which contains both two-fluid effects and the inflow pressure.

\begin{figure}[htbp]
\begin{center}
\ifjournal
\includegraphics[width={0.8\columnwidth},clip]{f10.eps}
\else
\includegraphics[width={0.8\columnwidth},clip]{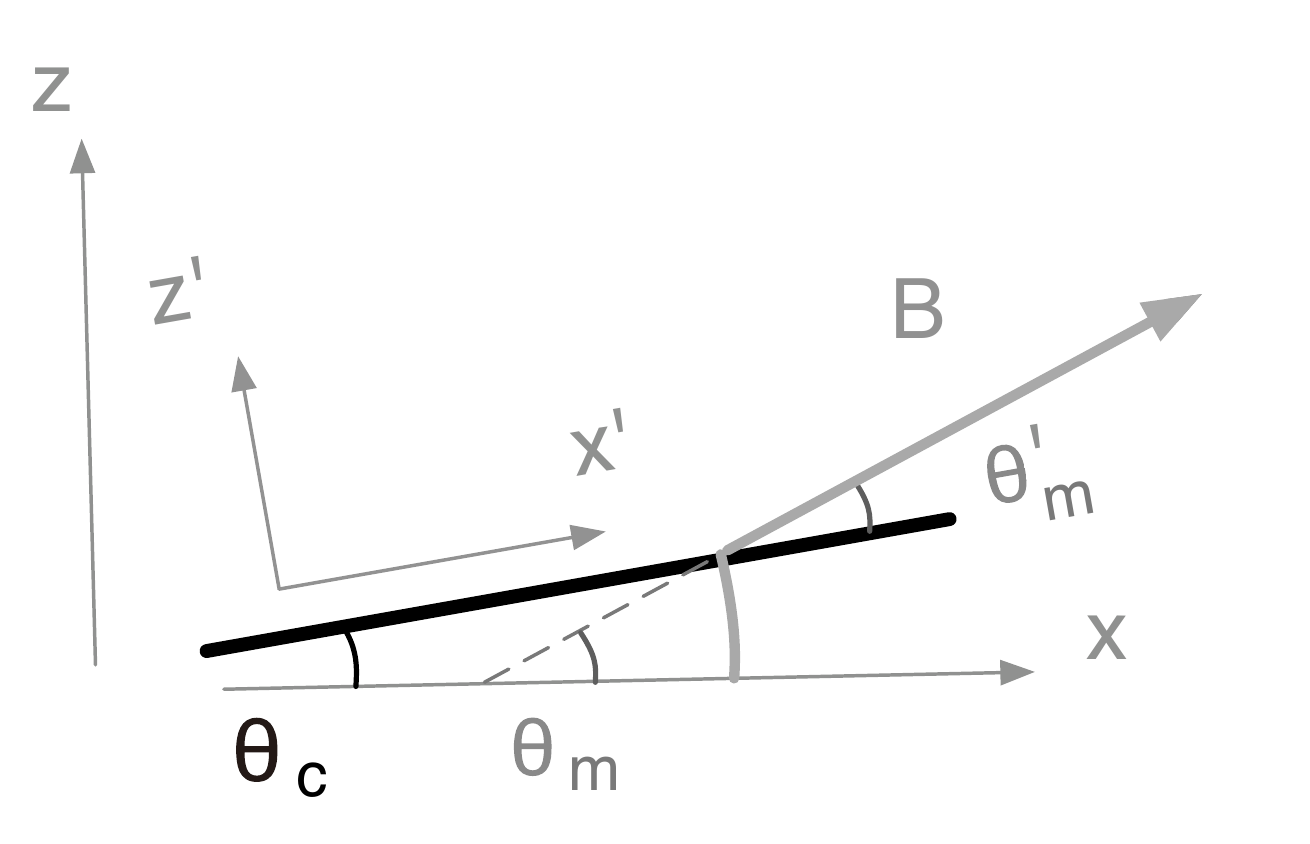}
\fi
\caption{Rotated coordinate for the Petschek current layer.
The thick black line stands for the current layer.
The two angles $\theta_c$ and $\theta_m$ are relevant to
the current layer and the upstream magnetic field line, respectively.
The angle $\theta'_m$ is the field line angle in the rotated frame.
\label{fig:slowshock}}
\end{center}
\end{figure}

Let us consider a rotated coordinate based on the Petschek current layer
(or the slow shock surface in \citet{lyu05}).
The new $x'z'$ coordinate is tilted from the simulation coordinate $xz$
by the angle of $\theta_c$ as shown in Figure \ref{fig:slowshock}.
The angles $\theta_m$ and $\theta'_m$ are the upstream field line angles
from the simulation frame and the rotated frame, respectively.
Since the electric field is almost uniform over these regions in our simulation,
we assume that the electric field $E_y$ is constant.
The relativistic stable conditions across the current layer are as follows:
\begin{eqnarray}
\label{eq:shock_eflux}
\Big[ 2\gamma w u_{z'} - \frac{c}{4\pi}E_yB_{x'}
\Big] = 0
\\
\label{eq:shock_pressure}
\Big[ \frac{2 w u_{z'}^2}{c^2} + 2p + \frac{B^2_{x'}}{8\pi}
\Big] = 0
\\
\label{eq:shock_momflux}
\Big[ \frac{2 w u_{x'}u_{z'} }{c^2} - \frac{B_{x'}B_{z'}}{4\pi}
\Big] = 0
\\
\label{eq:shock_frozen}
\Big[ v_{x'}B_{z'}-v_{z'}B_{x'} \Big] = 0
\\
\label{eq:shock_continuity}
\Big[ n u_{z'} \Big] = 0
\\
\label{eq:bn}
\Big[ B_{z'} \Big] = 0,
\end{eqnarray}
where the brackets stand for the jump condition in the $z'$ direction. 
We employ the assumption of $v_{x'u}=0$, $B_{x'd}=0$,
where $u$ and $d$ denote the upstream and the downstream properties.
We confirmed that these assumption are fair, especially $B_{x'd}=0$.
Then, equation \ref{eq:shock_frozen} yields
\begin{equation}
\label{eq:vB}
E_y = -\frac{v_{z'u}}{c}B_{x'u} = \frac{v_{x'd}}{c}B_{z'}
.
\end{equation}
From equations \ref{eq:shock_eflux} and \ref{eq:vB},
\begin{equation}
\label{eq:luyb1}
\Big( 2\gamma^2_u w_u+\frac{B^2_{x'u}}{4\pi} \Big) v_{z'u} =
2\gamma_d w_d u_{z'd}
.
\end{equation}
From equation \ref{eq:shock_momflux},
\begin{equation}
-\frac{B_{x'u}B_{z'}}{4\pi}
=
\frac{2w_d u_{x'd}u_{z'd}}{c^2}
.
\end{equation}
Eliminating $B_{z'}$ with equation \ref{eq:vB},
\begin{equation}
\label{eq:luyb2}
v_{z'u}\frac{B^2_{x'u}}{4\pi}
=
2\gamma_d w_d u_{z'd} \big(\frac{v_{x'd}}{c}\big)^2
.
\end{equation}
From equations \ref{eq:luyb1} and \ref{eq:luyb2},
we obtain
\begin{equation}
\big(\frac{v_{x'd}}{c}\big)^2
= \frac{B^2_{x'u}/4\pi}{2\gamma^2_u w_u+B^2_{x'u}/4\pi}
= \frac{\sigma_{\varepsilon,u} \cos^2\theta'_m}{1+\sigma_{\varepsilon,u} \cos^2\theta'_m}
,
\end{equation}
where we set $\sigma_{\varepsilon,u}=B^2_{u}/[4\pi(2\gamma^2_u w_u)]$.
We also obtain
\begin{equation}
\big(\frac{v_{z'u}}{c}\big)^2
= \frac{B^2_{z'}/4\pi}{2\gamma^2_u w_u+B^2_{x'u}/4\pi}
= \frac{\sigma_{\varepsilon,u} \sin^2\theta'_m}{1+\sigma_{\varepsilon,u} \cos^2\theta'_m}
.
\end{equation}
Then, we discuss the angles in the limit of $\sigma_{\varepsilon,u} \gg 1$.
Approximating $v_{z'u} = - c \tan \theta'_m$ and $w_d=4p_d$,
equations \ref{eq:shock_pressure} and \ref{eq:luyb1} can be modified as follows:
\begin{eqnarray}
\frac{B^2_u}{8\pi} \cos^2 \theta'_m &=& \frac{ 8\gamma^2_dp_dv^2_{z'd} }{c^2} + 2p_d \\
-c \frac{B^2_u}{4\pi} \cos \theta'_m \sin \theta'_m &=& 8\gamma^2_dp_dv_{z'd}
.
\end{eqnarray}
We immediately obtain
\begin{eqnarray}
p_d &=& \frac{B^2_u}{16\pi} \cos^2 \theta'_m \\
\gamma_d &=& \sqrt{\sigma_{\varepsilon,u}} \cos \theta'_m \\
v_{z'd} &=& - \frac{ c \tan \theta'_m }{ 2 \sigma_{\varepsilon,u} \cos^2 \theta'_m }
\end{eqnarray}
It is reasonable that
the outflow Lorentz factor is similar to
that of upstream Alfv\'{e}n speed, $\sqrt{1+\sigma_{\varepsilon,u}}$.
Considering that the outflow travels toward the $+x$ direction,
we find
\begin{eqnarray}
\label{eq:shockangle}
\theta_c \sim \theta'_m / ( 2 \sigma_{\varepsilon,u} ).
\end{eqnarray}
This means that the Petschek outflow channel becomes narrower and narrower,
as the upstream flow is more and more magnetically dominated.

\begin{figure}[htbp]
\begin{center}
\ifjournal
\includegraphics[width={0.9\columnwidth},clip]{f11.eps}
\else
\includegraphics[width={0.9\columnwidth},clip]{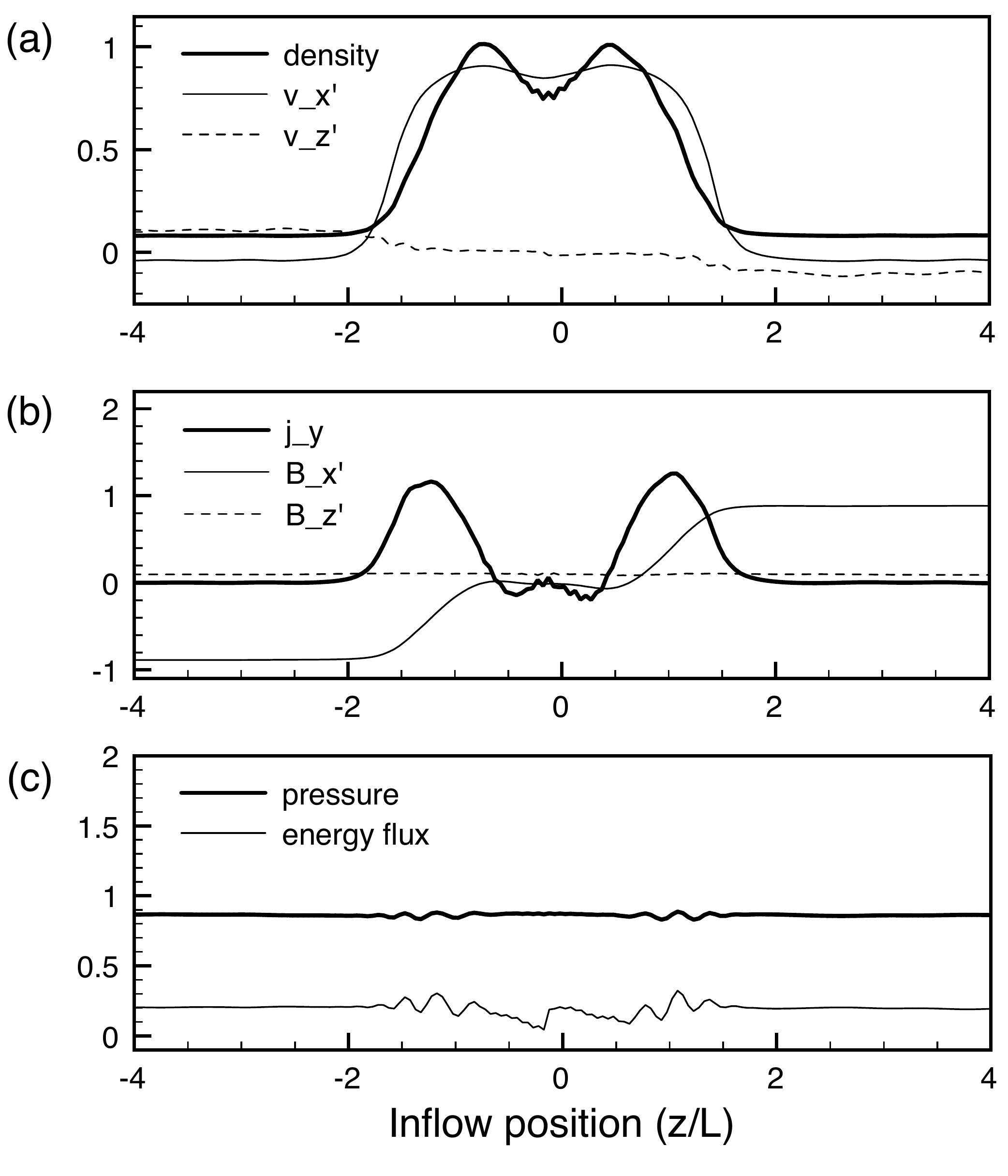}
\fi
\caption{Physical properties across the Petschek-type current layers
at $x/L=100$ at $t/\tau_c=400$.
(\textit{a}) Normalized plasma density $\gamma n / n_0$,
tangential plasma velocity $v_{x'}/c$,
normal plasma velocity $v_{z'}/c$,
(\textit{b}) the out-of-plane electric current $j_y/j_0$,
tangential magnetic field $B_{x'}/B_0$,
normal magnetic field $B_{z'}/B_0$ (eq. \ref{eq:bn}),
(\textit{c}) 
total pressure (eq. \ref{eq:shock_pressure}; normalized by $B^2_0/8\pi$),
and energy flow (eq. \ref{eq:shock_eflux}; normalized by $cB^2_0/8\pi$).
\label{fig:cut}}
\end{center}
\end{figure}
In our simulation,
we observe the Petschek-type current structures in runs L1-L5.
In the other runs, as discussed,
we could not solve the late-time evolution because of the numerical problems.
In runs L1-L5, we measured the angle of the Petschek current layers in the following way.
Near the current layer, we assume the inclined coordinate
assuming an arbitrary angle $\theta_c$ like Figure \ref{fig:slowshock}.
Then, across the current layer,
we look at the relativistic jump conditions across the $z'$ direction
(eqs. \ref{eq:shock_eflux}-\ref{eq:bn}).

Figure \ref{fig:cut} shows one example,
physical properties across the current layers
at $x/L=100$ at $t/\tau_c=400$,
as indicated by the white line in Figure \ref{fig:snap200}\textit{d}.
In this case, the oblique frame properties are calculated
by using an angle $\theta_c = 0.125$, and
the opposite rotation is applied to the properties of the lower half and the upper half.
We note that the neutral plane is slightly off-center ($z/L\sim -0.15$)
in this very late stage because of the open boundary conditions.
In the dense plasma region between the two current peaks,
we observe fast reconnection outflow $v_{x'}\sim 0.9c$ (Figure \ref{fig:cut}).
We also observe noises in the properties near the center and
the flux properties in the current layers; however,
we think that they are sufficient for the purpose of this study.

Varying $\theta_c$ with $\Delta \theta_c=0.025$,
we find out the best angle,
which minimizes the variation of the above variables.
Among them, the energy flux and the tangential momentum flux
(Equations. \ref{eq:shock_eflux} and \ref{eq:shock_momflux};
Figure \ref{fig:cut}\textit{c})
in the outflow region and in the current layers are very sensitive,
and so they give a reasonable estimate of $\theta_c$.
We can also confirmed that
the obtained angles are consistent with the topological structure,
because the distance between the current peaks is $\sim$2.4 and
the location is $x/L=100$.
We repeat this procedure at various points along the well-developed current layers,
where the structure is not influenced by the backward plasma flow around the plasmoids.
Repeating the analyses at various time steps,
we obtain the typical $\theta_c$ angle for the specific run.

\begin{figure}[htbp]
\begin{center}
\ifjournal
\includegraphics[width={\columnwidth},clip]{f12.eps}
\else
\includegraphics[width={\columnwidth},clip]{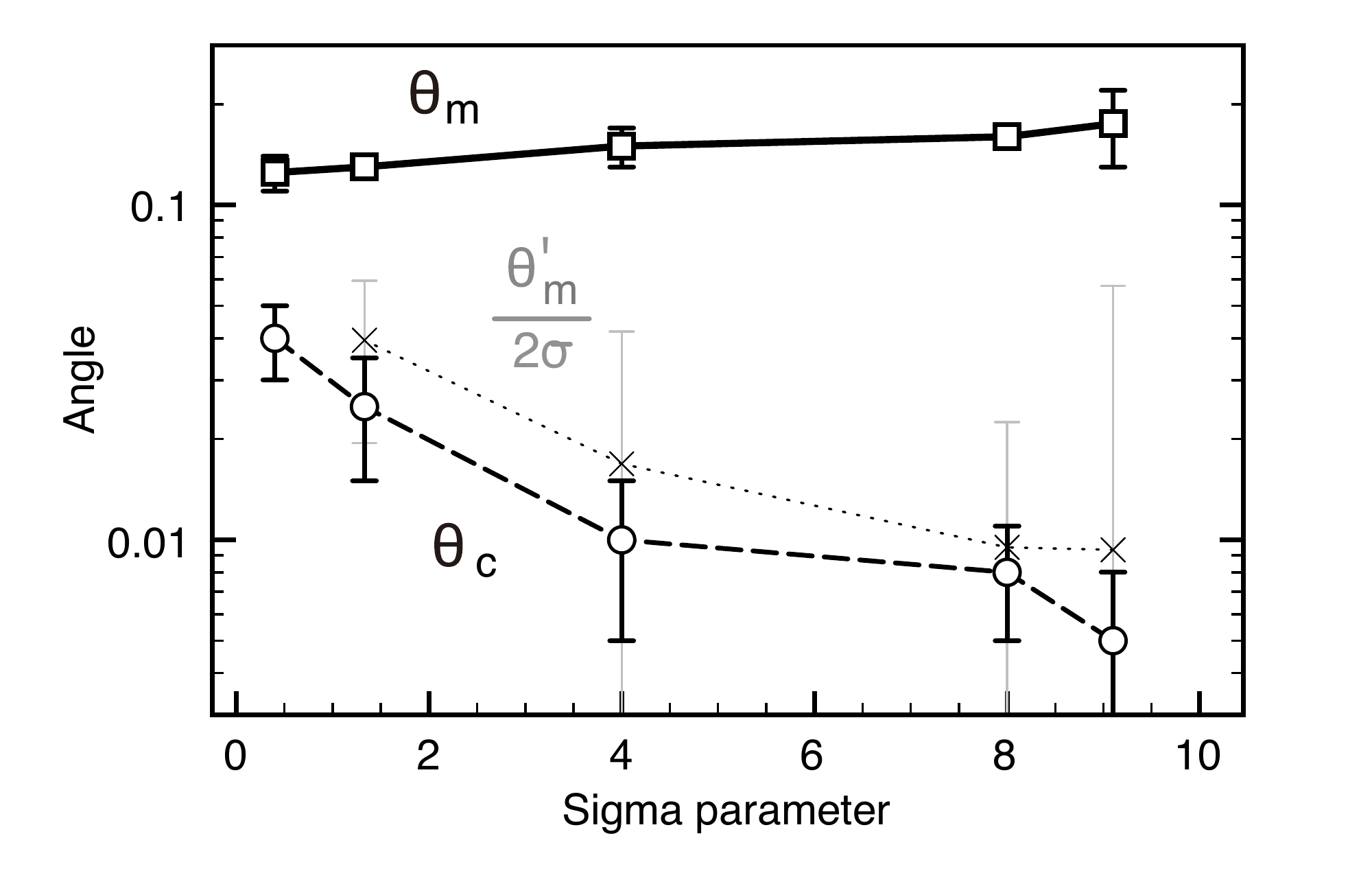}
\fi
\caption{Petschek angle analysis.
The magnetic field line angle $\theta_m$ (\textit{solid line}),
the current layer angle $\theta_c$ (\textit{dashed line}),
and
the estimated magnetic field line angle $\theta'_m/2\sigma_{\varepsilon,in}$
(\textit{dotted line})
are presented as a function of the initial parameter $\sigma_{\varepsilon,in}$.
\label{fig:lyubshock}}
\end{center}
\end{figure}

Figure \ref{fig:lyubshock} compares the obtained angles
by the above analysis in runs L1--L5.
The dashed line shows the current layer angle $\theta_c$.
The typical field line angle $\theta_m$ is also
measured in the upstream side of the current layers,
and they are presented in the solid line.
We find that the angle $\theta_c$ becomes narrower
as the inflow parameter $\sigma_{\varepsilon,in}$ increases.
On the other hand, the field line angle $\theta_m$ shows the opposite trend.
Considering that the reconnection rate increases
as $\sigma_{\varepsilon,in}$ increases,
it is quite reasonable that $\theta_m$ increases.
We expect that the angle is eventually asymptotic to $\theta_m \sim \pi/4$.
The dotted line in Figure \ref{fig:lyubshock} shows the theoretical angle.
It is estimated by substituting
$\sigma_{\varepsilon,u}\sim\sigma_{\varepsilon,in}$
in equation \ref{eq:shockangle}.
In the present parameter range,
we find an excellent agreement between two shock angles
$\theta_c$ and $\theta'_m/(2\sigma_{\varepsilon,in})$
(the dotted line in Figure \ref{fig:lyubshock}).
Although we discuss stable current layers in mildly relativistic runs,
we expect that the theory shows good agreement in the higher-$\sigma$ regime,
where the theory was originally designed.

\subsection{Uniform resistivity case\label{sec:uni}}

\begin{figure}[htbp]
\begin{center}
\ifjournal
\includegraphics[width={0.8\columnwidth},clip]{f13a.eps}
\includegraphics[width={0.8\columnwidth},clip]{f13b.eps}
\else
\includegraphics[width={0.8\columnwidth},clip]{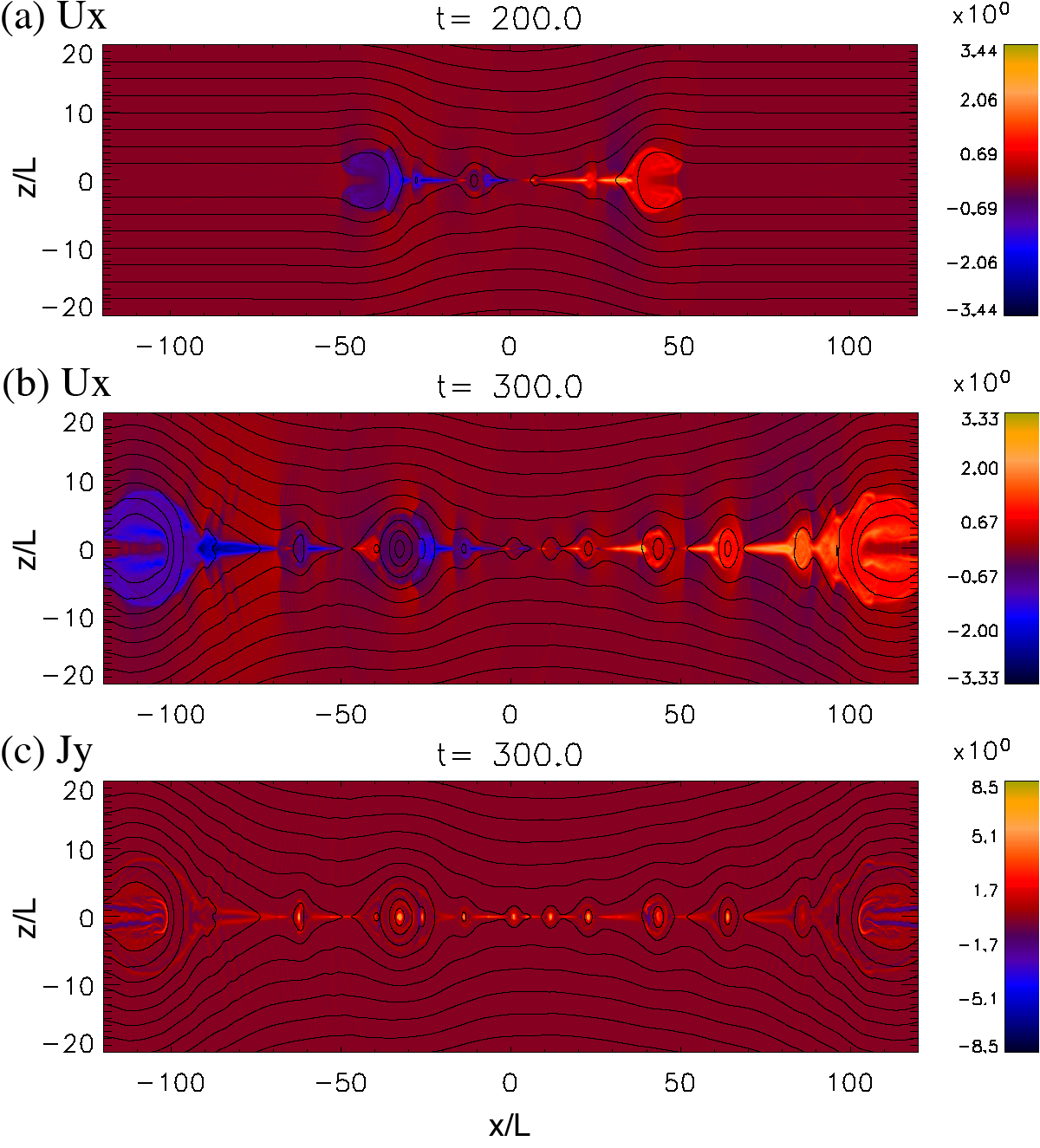}
\includegraphics[width={0.8\columnwidth},clip]{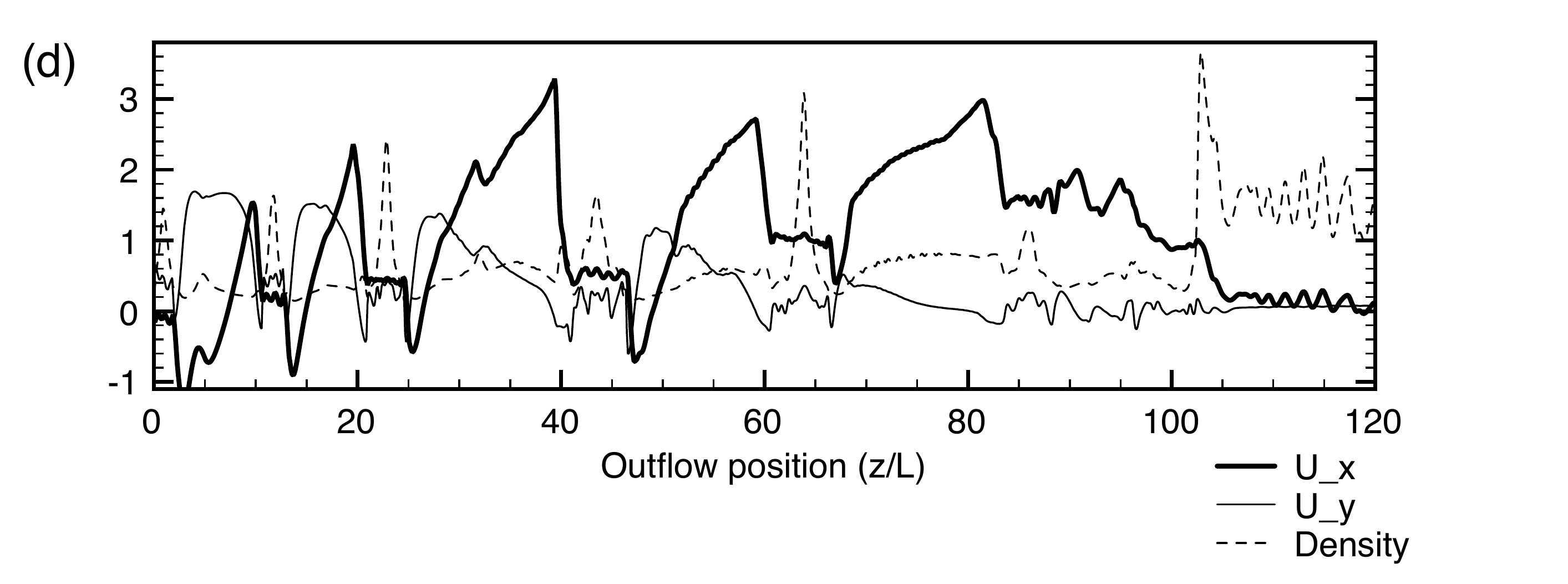}
\fi
\caption{
Large-scale snapshots of run U3:
(\textit{a}) the $x$-component of the plasma 4-velocity $u_x/c$ at $t/\tau_c=200$,
(\textit{b}) the $x$-component of the plasma 4-velocity $u_x/c$ at $t/\tau_c=300$, and
(\textit{c}) the out-of-plane current $j_y/j_0$ at $t/\tau_c=300$.
The black lines show magnetic field lines.
(\textit{d}) The outflow 4-velocity $u_x/c$,
the out-of-plane 4-velocity $u_y/c$,
and
the plasma density $\gamma n/n_0$ at $t/\tau_c=300$,
along the right half of the outflow line ($z=0$).
\label{fig:uni}}
\end{center}
\end{figure}

In order to study the role of the resistivity,
we also carried out another simulation run
with a uniform resistivity (run U3 in Table \ref{table}).
The parameters are the same as those of run L3,
but the resistivity is uniformly set.
Its effective Reynolds number is $R_M=3000$.
Compared with run L3,
the system evolves slower
primary due to the low resistivity at the reconnecting $X$-point.
The top three panels in Figure \ref{fig:uni} present
the late-time evolution of run U3,
at $t/\tau_c=200$ and $300$.
Figure \ref{fig:uni}\textit{d} shows
the properties along the outflow line at $t/\tau_c=300$.
At $t/\tau_c=200$, the reconnection outflow is
still only half way to the boundaries.
The reconnecting current sheet contains several secondary structures.
We think that this is due to the slower evolution of the system.
There is sufficient time for secondary structures to grow.
The biggest plasmoids reach the boundaries around $t/\tau_c=300$.
Now we observe a formation of multiple big islands
inside the reconnecting current sheet.
As wee see in the profiles in Figure \ref{fig:uni}\textit{d},
multiple magnetic reconnections take place and expel outflows
between these islands.
The outflow 4-velocity reaches $u_x/c \sim 3$ at various local points,
and the global flow speed seems to be $u_x/c \sim 1$--$2$.
The density spikes in Figure \ref{fig:uni}\textit{d}
are identical to the $O$-points,
magnetic nulls at the center of plasmoids.
Although the out-of-plane flow is very small, $u_y/c \ll 1$,
these high-density plasmas carry the electric current
inside the $O$-type regions (Figure \ref{fig:uni}\textit{c}).
On the other hand, around several regions between the islands,
we see that the out-of-plane 4-velocity is enhanced,
$u_y/c \sim 1$ or $1.5$ (Figure \ref{fig:uni}\textit{d}).
They are related to thin current sheets between plasmoid islands.
The plasma temperature is typically $p/nmc^2 \sim 2$ in the outflow region,
and it becomes very high $p/nmc^2 \sim 4$--$5$ around the $O$-points.
The simulation continues until $t/\tau_c\sim 345$
shortly after the plasmoids completely went through the boundaries.

When the plasmoid islands appear,
its typical timescale seems to be tens of $\tau_c$,
and it is faster than an estimated
timescale of the resistive tearing mode,
$R_M^{3/5}$ or $hR_M^{3/5} \sim \mathcal{O}(10^2)$.
We think that
the island formation is enhanced by the two-fluid effect,
which was introduced in our simulation.
Since our Ohm's law (eq. \ref{eq:ohm}) contains
the fluid inertial term $\partial_t (h_p u_{py})$,
the tearing mode can grow more explosively than
the classical resistive MHD case.
If we use the specific condition of $u_y\sim 1$,
the timescale of the relativistic collisionless tearing mode \citep{zelenyi79}
is $\mathcal{O}(10)$, too.

Although the system evolution is slower than the reference run L3,
we note that the reconnection still remains fast,
at least during this simulation run,
and it may be related to island formation.
If we discuss the global structure
by filtering out the local plasmoid islands,
the average plasma inflow speed is $v_y \sim 0.1c$, and
the reconnection electric field is $E_y \sim 0.1 B_0$.
This may be an interesting hint to discuss
the problem of a fast magnetic reconnection.

\section{DISCUSSION}

First, let us briefly compare our results
with the one-fluid work by \citet{naoyuki06} (here referred as WY06).
They employ an relativistic Ohm's law
\begin{equation}
\label{eq:relaohm}
\vec{E}+\frac{\vec{v}}{c}\times\vec{B}=(\eta/\gamma')\vec{j},
\end{equation}
where $\gamma'$ is the Lorentz factor of the one-fluid MHD motion.
Our Ohm's equation (eq. \ref{eq:ohm}) differs in the following two ways.
First, since our equation contains the fluid inertial term,
by definition our model describes better physics. 
Second, we do not consider the factor of $1/\gamma'$.
However, we consider finite resistivity only near the $X$-point,
where $\gamma'$ is close to the unity.
The fastest run in WY06 is directly equivalent to
our reference run L3.
The reconnection geometry looks similar.
However, we find various minor differences.
The maximum outflow 4-velocity is $u_x/c\sim 2.2$--$2.3$ in WY06,
while we often observe faster value ($u_x/c>3$)
(e.g., Figure \ref{fig:outflow}\textit{b}).
This is quite probably due to the two-fluid effect and the grid condition.
Since we also deal with the out-of-plane motion $u_y$,
the Lorentz factor can be larger even when it contains a contribution from $v_y$.
Also, WY06 employed nonuniform grids,
and then physical quantities in the distant outflow region
are often averaged in the larger computational cells.
Regarding the structure of the Petschek-type reconnection,
WY06 implied that
their shock angle becomes narrower in the relativistic regime
like \citet{lyu05} predicted.
Taking the inflow pressure into account,
we clarified that
a modified \citet{lyu05} theory well explain
the simulation results (Figure \ref{fig:lyubshock}).
Our angle is narrower than that of WY06.
This is probably because the isotropic plasma pressure is
usually overemphasized in one-fluid model,
and
because the amplitude of the effective resistivity may be different. 

We obtained several implications for
relativistic reconnection models.
Although we do not obtain
a steady state Sweet--Parker reconnection,
in all our runs, the early evolutions of main reconnection runs
will be good hints to understand relativistic Sweet--Parker reconnection.
For example, in run L3, along the outflow line,
we found that plasma temperature becomes very high
(e.g., Figure \ref{fig:outflow}\textit{e}).
Consequently, the relativistic enthalpy substantially increases,
$\Delta w \sim 4n \Delta T > nmc^2$ in the outflow region.
Therefore, the relativistic reconnection model cannot neglect
the relativistic gas pressure in the outflow region \citep{lyu05,zeni08c}. 
In the Petschek reconnection regime,
we observed that the bifurcated current layers and
their angle becomes narrower and narrower
as the inflow becomes more and more magnetically dominated.
With minor modification, this trend is
well consistent with the argument proposed by \citet{lyu05}.

The ultrarelativistic limit of $\sigma_{\varepsilon,in} \gg 1$ is
of strong astrophysical interests.
Our parameter study suggested that
the reconnection rate is asymptotic to
its upper limit of $\sim$1 in that regime.
Indeed, in the most extreme case (run L9),
we observe fast reconnection with super fast inflow.
We expect that reconnection is super fast in the high-$\sigma$ regime.
Such a fast reconnection rate implies that
the separatrix angle will be
wide open---asymptotic to $45^\circ$ in the steady stage. 
In fact, our mildly relativistic runs show
the magnetic field line angle $\theta_m$ constantly increases
(e.g., Figure \ref{fig:lyubshock}) in the Petschek-type steady regime.
In a sense this is reasonable,
because there are less current carrier in such a regime.
When the separatrix becomes open,
the field reversal current for the reconnected fields $\pm B_z$
partially cancel the field reversal current for the antiparallel fields $\pm B_x$,
therefore the system needs less electric current.
We do not know whether or not the bifurcated Petschek-type solution exists 
in the high-$\sigma$ regime.
Since the current layer becomes too flat
and the central diffusion region tends to expand,
the reconnection current sheet may remain
in a single thick current layer for a long time.

We think that an important feature of the reconnection in the high-$\sigma$ regime
is the shortage of the current carrier.
As the authors discussed through
PIC simulation and the two-fluid theory \citep{zeni08c},
when reconnection environment is magnetically dominated and
runs out of current carriers,
the displacement current induces the strong electric field.
It leads to a faster reconnection rate and
the expansion of the central Sweet--Parker region.
From MHD viewpoint,
it means the enhancement of the effective resistivity;
however, we note that
the conventional single-fluid RMHD simulations have
no explicit upper limit of plasma currents.
Due to the enhancement of the reconnection field,
we find a large electric-dominated region in run L9,
where the field is electrically dominated, $(E^2-B^2)>0$,
around the $X$-type region. 
Plasmas are no longer magnetized there, and then
powerful DC acceleration will occur \citep{zeni01}.
Therefore, we expect that the high-$\sigma$ reconnection
is a favorable source of nonthermal particles acceleration.
Long-term evolution, theoretical modeling,
and particle acceleration
in the high-$\sigma$ regime will be left for future work.

In this work, we mainly use
the energy-based magnetization parameter $\sigma_{\varepsilon,in}$,
because it seems to be a better measure of reconnection property
than the conventional magnetization parameter $\sigma_{m,in}$.
In fact, as long as we surveyed
(Sections \ref{sec:cases} and \ref{sec:slowshock}),
run L4 with lower pressure resembles run L5 rather than run L3.
However, we still observe minor differences, and
the running out of the current carrier should be controlled by $\sigma_{m}$.
So, we conclude that both two magnetization parameters
$\sigma_{\varepsilon,in}$ and $\sigma_{m,in}$ are important.

We also demonstrated that
the spatial profile of the resistivity has great influence
on the system evolution. 
As recognized in many works in the nonrelativistic regime,
the spatially localized resistivity
leads to the Petschek-like reconnection
with bifurcated current layers (e.g., \citet{ugai77,scho89}).
On the other hand, the uniform resistivity case
exhibits a single current layer with secondary islands.
Considering that the governing equations are almost the same
outside the localized resistivity point,
it is impressive to see such a contrast.
The dependence on the amplitude of the resistivity $\eta_{eff}$,
the spatial profile, and the other physical models,
will be left for future work.
Regarding the outflow structure,
at present PIC simulations of the relativistic magnetic reconnection
exhibit a laminar outflow
without islands or with small minor islands in a main reconnection region
\citep{zeni01,zeni07,claus04,zeni08b,zeni08c}. 
On the other hand, in the nonrelativistic regime,
\citet{dau07} demonstrated a very interesting result
by using a large-scale PIC simulation.
They showed that reconnection outflow is highly influenced by
a continuous formation of secondary islands,
and its global picture looks similar to
our uniform resistivity run (Figure \ref{fig:uni}\textit{b}).
We do not know whether the relativistic reconnection is influenced by
such continuous island formation,
it is worth investigating by using a larger relativistic PIC simulation.
We also observe small islands in higher-$\sigma$ runs,
and so the island formation may also be controlled by
the upstream parameters ($\sigma_{m,in}$ and $\sigma_{\varepsilon,in}$).



On the viewpoint of numerical accuracy,
we confirmed that our primitive variable solver is sufficiently accurate.
In Appendix A (Figure \ref{fig:error}),
the numerical error of our solver is presented.
In our range of interest ($|u|/c \lesssim 10^1$ and $p/nmc^2 \lesssim 10$),
the relative errors
in the restored primitive variables are very small,
$\sim \mathcal{O}(10^{-15})$.
Therefore, the worst estimate of
the accumulated error would be still negligible,
$10^{1} \times 10^{-14} \cdot (400\tau_c/\Delta t ) < 10^{-8}$.
On the other hand,
in order to further explore the higher-$\sigma$ conditions,
we have to improve the numerical scheme.
At present, the modified Lax--Wendroff scheme seems to be the bottle neck.
It is not ideal to describe shocks,
while the discontinuities around the magnetic pileup regions are
always difficult to solve in a nonsteady stage of reconnection.
We plan to employ a more stable scheme
such as HLL schemes \citep{mizuno06, mig09}
in order to study long-term evolution in the high-$\sigma$ regime.


Finally, let us discuss potential targets beyond this work.
A straightforward extension will be magnetic reconnection
with the out-of-plane magnetic field ($B_y$ or the ``guide field'').
In is already known that
PIC simulations with the guide field
exhibit global charge separation in a reconnection region,
and so the neutral one-fluid approximation already breaks down \citep{zeni08}.
Therefore we have to solve the positron and electron evolution separately
without the symmetric assumption. 
In the Petschek regime, \citet{lyu05} claimed that
(1) the relativistic magnetic reconnection involves
rotational discontinuities as well as slow shocks,
and that (2) the compressed guide field flux inside the outflow channel is
the main energy carrier in the magnetically dominated regime.
These properties are worth checking in future simulations.
In three dimensions,
it is known that the reconnection current sheet is unstable to
the relativistic drift kink instability,
which arises from the counter-streaming two-fluid motion of
positron fluids and electron fluids \citep{zeni05a,zeni07,prit96,dau99}.
These three-dimensional evolutions should be
carefully compared with PIC simulations, so that
we can study the larger problems
such as magnetar flares and global pulsar magnetospheres
by using a relativistic two-fluid model.
Of course, it is critically important to establish
an improved (theoretical or empirical) resistivity model,
which highly affects the system evolution.

\section{SUMMARY}

We carried out relativistic two-fluid MHD simulation
of a magnetic reconnection in an electron--positron pair plasma.
The interspecies friction term works as an effective resistivity,
and then we successfully demonstrated
the large-scale evolution of relativistic magnetic reconnection.
The system evolves from
a Sweet--Parker-like fast reconnection
to a Petschek-like reconnection with bifurcated current layers.
Open boundary conditions enable us to observe long-term evolutions,
and we find a Petschek structure, which is quite stable.
As \citet{lyu05} predicted,
the current layer angle becomes substantially narrower
when the reconnection inflow is more magnetically dominated.
Meanwhile, we find that
the reconnection rate goes up to $\sim$1 in extreme cases,
which implies that efficient particle acceleration occurs
in the electric-dominated region.
In addition, we demonstrate that the system evolution is 
controlled by the resistivity model.
We emphasize that
the large-scale reconnection problems are
investigated with a two-fluid RMHD model.
Beyond the single-fluid RMHD approximation,
multifluid models will be good alternatives to study
astrophysical plasma problems
which involve magnetic dissipation.



\begin{acknowledgments}
The authors are grateful to N. Watanabe, R. Yoshitake, and
M. Kuznetsova for helpful comments.
The authors thank the anonymous referee for his/her constructive comments
which helped them to improve the manuscript. 
This research was supported by the NASA Center for Computational Sciences,
and NASA's \textit{MMS} SMART mission.
S. Z. gratefully acknowledges
support from NASA's postdoctoral program.
\end{acknowledgments}


\begin{table}[htbp]
\caption{List of Simulation runs}\label{table}
\begin{tabular}{ccccccccccl}
Name & Domain Size & Grid Points & $c\Delta t/\Delta_g$ & $n_{in}/n_0$ & $p_{in}/p_0$ &
$\sigma_{m,in}$ & $\sigma_{\varepsilon,in}$ & $c_{A,in}/c$ \\
\hline
S3 & 80 $\times$ 40 & 1200 $\times$ 600 & 0.3 & 0.1 & 1.0 & 20 & 4 & $0.894$ \\
M3 & 240 $\times$ 120 & 3600 $\times$ 1800 & 0.3 & 0.1 & 1.0 & 20 & 4 & $0.894$ \\
L1 & 240 $\times$ 120 & 4800 $\times$ 2400 & 0.2 & 1.0 & 1.0 & 2 & 0.4 & $0.535$ \\
L2 & 240 $\times$ 120 & 4800 $\times$ 2400 & 0.2 & 0.3 & 1.0 & 6.67 & 1.33 & $0.816$ \\
L3 & 240 $\times$ 120 & 4800 $\times$ 2400 & 0.2 & 0.1 & 1.0 & 20 & 4 & $0.894$ \\
L4 & 240 $\times$ 120 & 4800 $\times$ 2400 & 0.2 & 0.1 & 0.3 & 20 & 9.1 & $0.953$ \\
L5 & 240 $\times$ 120 & 4800 $\times$ 2400 & 0.2 & 0.05 & 1.0 & 40 & 8 & $0.949$ \\
L6 & 240 $\times$ 120 & 4800 $\times$ 2400 & 0.2 & 0.03 & 1.0 & 66.7 & 13.3 & $0.964$ \\
L7 & 240 $\times$ 120 & 4800 $\times$ 2400 & 0.2 & 0.02 & 1.0 & 100 & 20 & $0.976$ \\
L8 & 240 $\times$ 120 & 4800 $\times$ 2400 & 0.2 & 0.01 & 1.0 & 200 & 40 & $0.988$ \\
L9 & 240 $\times$ 120 & 4800 $\times$ 2400 & 0.2 & 0.005 & 1.0 & 400 & 80 & $0.994$ \\
U3 & 240 $\times$ 120 & 4800 $\times$ 2400 & 0.2 & 0.1 & 1.0 & 20 & 4 & $0.894$ &
*uniform resistivity \\
XL3 & 240 $\times$ 120 & 9600 $\times$ 4800 & 0.2 & 0.1 & 1.0 & 20 & 4 & $0.894$ \\
\end{tabular}
\end{table}

\appendix

\section{Obtaining primitive variables}

The rest energy density $D$,
the momentum density $\vec{m}$,
and the energy density $\mathcal{E}$
are related to the primitive variables in the following way:
\begin{eqnarray}
D &=& \gamma n mc^2 \\
\vec{m} &=& [ \gamma (e+p) \vec{u} ] /{c^2} \label{mom} \\
\mathcal{E} &=& \gamma^2 (e+p) - p \label{ene}.
\end{eqnarray}
For convenience, we introduce $M=|\vec{m}|c$ and $\bar{u}=|\vec{u}|/c$.
We consider the case of $M > 0$,
because we immediately know $\bar{u}=0$ when $M = 0$.
Approximating the enthalpy $(e+p) = n mc^2 + G p$
by $G=\Gamma/(\Gamma-1)$,
equations \ref{mom} and \ref{ene} become
\begin{eqnarray}
M &=& \gamma(nmc^2+Gp)\bar{u} = ( D + \gamma G p ) \bar{u} \label{mom2} \\
\mathcal{E} &=& \gamma^2 (nmc^2+Gp) - p = \gamma D + (\gamma^2 G-1) p \label{ene2}
\end{eqnarray}
From equations \ref{mom2} and \ref{ene2},
we can eliminate $p$ in the following way:
\begin{equation}
(\gamma^2G-1) M - \gamma G \bar{u} \mathcal{E} = (\gamma^2G-1)D \bar{u} - \gamma^2 G \bar{u} D
\end{equation}
\begin{equation}
\gamma G \bar{u} \mathcal{E} = (\gamma^2G-1) M + D\bar{u} \label{eq:appendix}
.
\end{equation}
Squaring equation \ref{eq:appendix} and
substituting $\gamma^2=1+\bar{u}^2$,
we obtain
\begin{eqnarray}
G^2 (\mathcal{E}^2-M^2) \bar{u}^4 - 2GMD \bar{u}^3 +
\Big[ G^2 \mathcal{E}^2 - 2GM^2(G-1) - D^2 \Big] \bar{u}^2
- \Big[ 2D(G-1)M \Big] \bar{u}
- \Big[ (G-1)M \Big]^2 = 0 \label{eq1}
\end{eqnarray}
We solve this equation to obtain the physically valid solution.
Other primitive variables are easily obtained
by using the solution $\bar{u}$. 
When the first coefficient $(\mathcal{E}^2-M^2)$ is negative
we immediately stop the simulation,
because such situation is physically invalid. 
We also checked the other conditions $D>0$ and $(\mathcal{E}-D)>0$.

\begin{figure}[htbp]
\begin{center}
\ifjournal
\includegraphics[width={0.45\columnwidth},clip]{f14a.eps}
\includegraphics[width={0.45\columnwidth},clip]{f14b.eps}
\else
\includegraphics[width={0.45\columnwidth},clip]{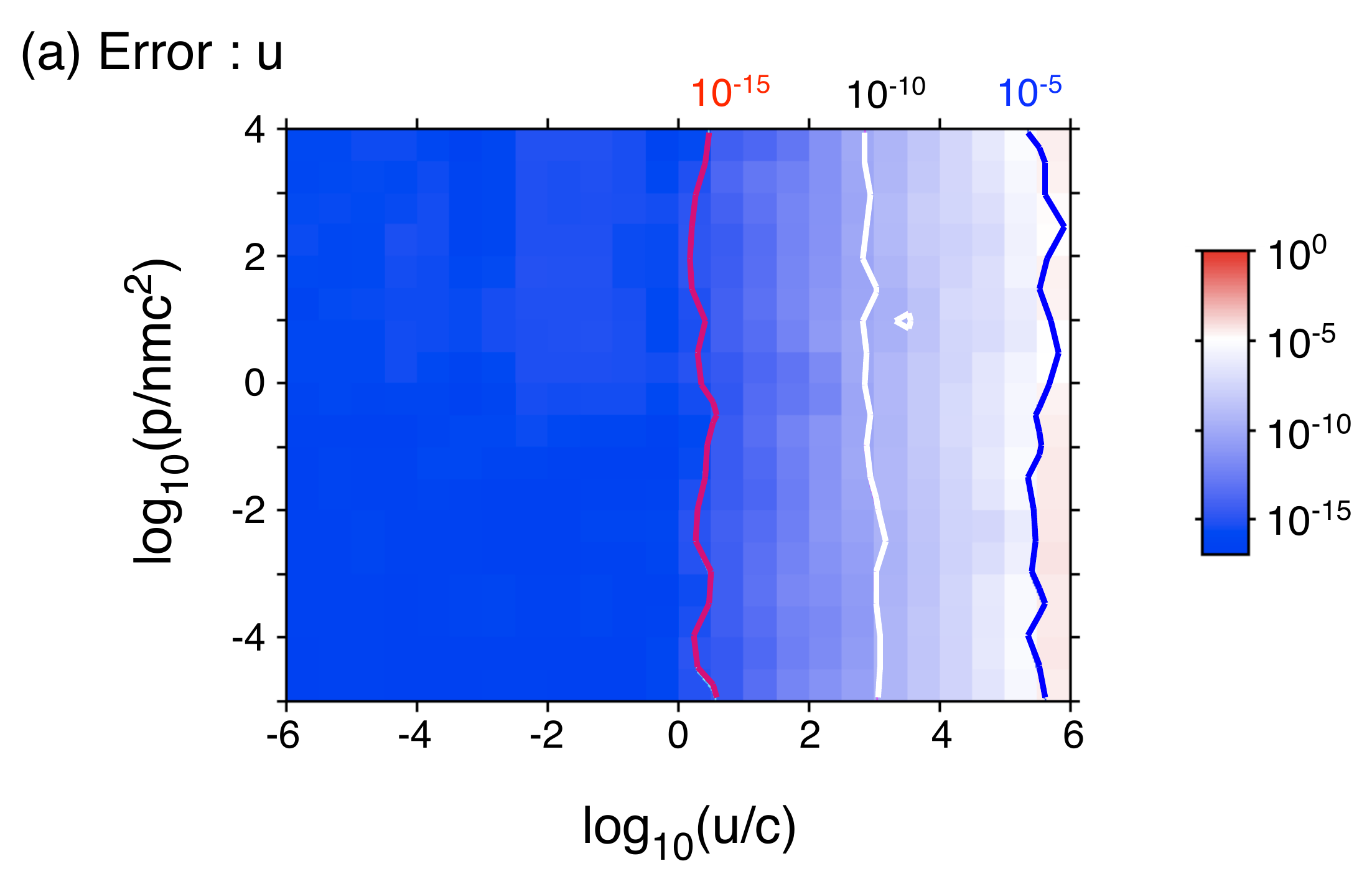}
\includegraphics[width={0.45\columnwidth},clip]{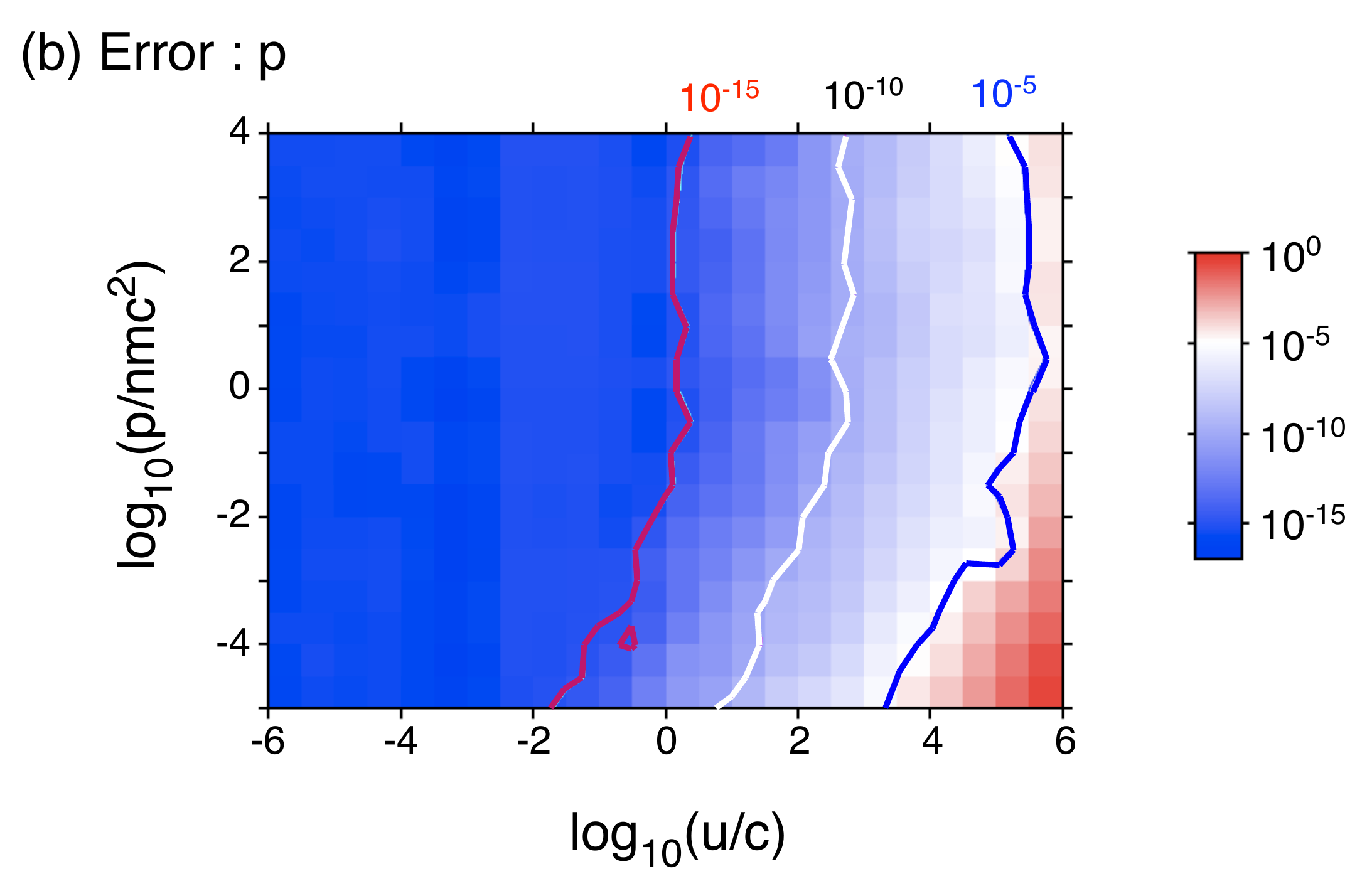}
\fi
\caption{
Numerical accuracy of our primitive variable solver.
Relative errors in the restored
(\textit{a}) $\bar{u}$ and (\textit{b}) $p$ are presented
as a function of $\bar{u}$ and $p/nmc^2$.
\label{fig:error}}
\end{center}
\end{figure}

The behavior of our primitive variable solver is
characterized by two parameters,
the relativistic bulk flow $\bar{u}$ and
the relativistic temperature $p/nmc^2$.
We benchmarked the numerical accuracy of our solver, and
Figure \ref{fig:error} shows the results as a function of the two parameters.
We can see that the quartic solution $\bar{u}$ is accurate
even in the ultrarelativistic regime of $\bar{u}\sim 10^{5}$.
The pressure $p$ is least reliable
in the limit of $p \ll nmc^2$ and $\bar{u} \gg 1$.
This is because the pressure is enclosed in the enthalpy term
$w=e+p=n [mc^2+G (p/nmc^2) ]$, but
the fluid macro properties are insensitive to $p$ in such cases.
The error in $n$ shows the same trend as that of $\bar{u}$.

\section{Brown Method}

We solve the quartic equations by using a simplified version of
the Brown method \citep{nuno03}.
Consider the following quartic equation:
\begin{eqnarray}
u^4 + c_3 u^3 + c_2 u^2 + c_1 u + c_0 = 0  \label{eq2}
\end{eqnarray}
where $c_n$ is the $n$th-order coefficient.
When we apply eq. \ref{eq1} to eq. \ref{eq2},
we find that the coefficients $c_n$ are all real,
and that $ c_0 < 0$.
In addition to two solutions ($u_1,u_2$),
eq. \ref{eq2} has one real negative solution ($u_3<0$) and
one real positive solution ($0<u_4$).
The first two ($u_1,u_2$) are usually complex conjugates, and
the real positive one $u_4$ is the physically valid solution
that we are looking for.
Let us consider the following two equations:
\begin{eqnarray}
(u-u_1)(u-u_2) &=& u^2+a_1u+a_0
\label{eq:a2}
\\
(u-u_3)(u-u_4) &=& u^2+b_1u+b_0
\label{eq:b2}
.
\end{eqnarray}
Examining the properties of $b_n$ and $c_n$,
we know that $a_n \in \mathbb{R}$ and
that $a_0>0$ and $b_0<0$.
Therefore the quartic equation (eq. \ref{eq1}) can be
decomposed into two quadratic equations with real coefficients. 
The relations between $a_n$, $b_n$, and $c_n$ are as follows:
\begin{equation}
\left\{ \begin{array}{lll}
c_3 &=& a_1 + b_1 \\
c_2 &=& a_1 b_1 + a_0 + b_0 \\
c_1 &=& a_0 b_1 + a_1 b_0 \\
c_0 &=& a_0 b_0
.\\
\end{array} \right.
\end{equation}
We further define $v := (a_1 - b_1), w := (a_0 + b_0), x := (a_0 - b_0)$.
Using $v, w, x$, we rewrite these equations
\begin{equation}
\label{eq:brown}
\left\{ \begin{array}{lll}
4 c_2 &=& c_3^2 - v^2 + 4 w \\
4 c_1 &=& 2c_3w - 2xv \\
4 c_0 &=& w^2 - x^2
.\\
\end{array} \right.
\end{equation}
We obtain the following relation:
\begin{eqnarray}
x^2v^2 = (c_3w-2c_1)^2=(w^2-4c_0)(4w-4c_2+c_3^2)
\end{eqnarray}
and then
\begin{eqnarray}
\label{eq:cubic}
w^3 - c_2 w^2 + (c_1c_3-4c_0) w + [ c_0(4c_2-c_3^2)-c_1^2 ]  = 0
.
\end{eqnarray}
We solve this third-order equation to obtain $w$,
paying attention to the numerical accuracy \citep{nuno96}.
Usually, we obtain two complex solutions and the one real solution
in Equation \ref{eq:cubic}, and so we employ the sole real solution $w$.
When we find three real solutions in Equation \ref{eq:cubic},
we may have multiple choices for $u$, because
Equation \ref{eq:a2} also has two real solutions instead of complex conjugates.

By using this $w$, we obtain $x = +\sqrt{w^2-4c_0}$.
Here we choose a positive square root, because we know that $a_0>0, b_0<0$.
The last variable $v$ can be obtained from eq. \ref{eq:brown} accordingly.
By using $w$, $x$, $v$, and $c_3$, we obtain $a_n$ and $b_n$,
and then we obtain $u_4$ as a positive root of eq. \ref{eq:b2}.

If we employed the standard Ferrari's method to solve the quartic equations,
we should use Cardano's transformation, $\bar{u} := u' - c_3/4$,
in order to eliminate the third-order coefficient $c_3$.
Then, the inverse transformation of $\bar{u}$ often causes
the cancellation of significant digits
when the solution is very small, $\bar{u} \sim 0$.
On the other hand, by using Brown method,
we are not so influenced by the cancellation of significant digits
when the solution is very small, $\bar{u}\sim 0$.


\begin{thebibliography}{}

\bibitem[Aschwanden(2006)]{aschwanden} M. J. Aschwanden 2006, Physics of the Solar Corona: An Introduction with Problems and Solutions (2nd Edition; Berlin: Springer), Chapter 10
\bibitem[Birk et al.(2001)]{birk01} G. T. Birk, A. R. {Crusius-W{\"a}tzel} and H. {Lesch} 2001, \apj, 559, 96
\bibitem[Bessho \& Bhattacharjee(2007)]{bessho07} N. Bessho and A. Bhattacharjee 2007, \pop, 14, 056503
\bibitem[Blackman \& Field(1993)]{bf93} E. G. Blackman and G. B. Field 1993, \prl, 71, 3481
\bibitem[Blackman \& Field(1994)]{bf94b} E. G. Blackman and G. B. Field 1994, \prl, 72, 494
\bibitem[Bucciantini et al.(2006)]{buc06} N. Bucciantini, T. A. Thompson, J. Arons, E. Quataert and L. Del Zanna 2006, \mnras, 368, 1717
\bibitem[Chen \& Shibata(2000)]{chen00} P. F. Chen and K. Shibata 2000, \apj, 545, 524
\bibitem[Clare \& Strottman(1986)]{clare86} R. B. Clare and D. Strottman 1986, Physics Reports, 141, 177
\bibitem[Coroniti(1990)]{coro90} F. V. Coroniti 1990, \apj, 349, 538
\bibitem[Daughton(1999)]{dau99} W. Daughton 1999, \pop, 6, 1329
\bibitem[Daughton \& Karimabadi(2007)]{dau07} W. Daughton and H. Karimabadi 2007, \pop, 14, 072303
\bibitem[Del Zanna et al.(2003)]{dz03} L. Del Zanna, N. Bucciantini and P. Londrillo 2003, \aap 400, 397
\bibitem[di Matteo(1998)]{dimatteo} T.~{di Matteo} 1998, \mnras, 299, L15
\bibitem[Drenkhahn(2002)]{dr02} G.~{Drenkhahn} 2002, \aap, 387, 714
\bibitem[Drenkhahn \& Spruit(2002)]{drs02} G.~{Drenkhahn} and H.~C. {Spruit} 2002, \aap, 391, 1141
\bibitem[Duncan \& Hughes(1994)]{duncan94} G. C. Duncan and P. A. Hughes 1994, \apj, 436, 119
\bibitem[Duncan \& Thompson(1992)]{duncan92} R. C. Duncan and C. Thompson 1992, \apj, 392, L9
\bibitem[Gammie et al.(2003)]{gammie03} C. F. Gammie, J. C. McKinney and G. T\'{o}th 2003, \apj, 589, 444
\bibitem[Hesse et al.(1999)]{hesse99} M. Hesse, K. Schindler, J. Birn and M. Kuznetsova 1999, \pop, 6, 1781
\bibitem[Hesse \& Zenitani(2007)]{hesse07} M. Hesse and S. Zenitani 2007, \pop, 14, 112102
\bibitem[Jaroschek et al.(2004)]{claus04} C. H. Jaroschek, R. A. Treumann, H. Lesch and M. Scholer 2004, \pop, 11, 1151
\bibitem[Karlick{\'y}(2008)]{karl08} M. {Karlick{\'y}} 2008, \apj, 674, 1211
\bibitem[Kirk \& Skj\ae raasen(2003)]{kirk03} J. G. Kirk and O. Skj\ae raasen 2003, \apj, 591, 366
\bibitem[Koide \& Arai(2008)]{koide08} S. Koide and K. Arai 2008, \apj, 682, 1124
\bibitem[Koide et al.(1996)]{koide96} S. Koide, K. Nishikawa and R. L. Mutel 1996, \apj, 463, L71
\bibitem[Koide et al.(1999)]{koide99} S. Koide, K. Shibata, and T. Kudoh 1999, \apj, 525, 727
\bibitem[Komissarov(1999)]{kom99} S. S. Komissarov 1999, \mnras, 303, 343
\bibitem[Komissarov(2006)]{kom06} S. S. Komissarov 2006, \mnras, 367, 19
\bibitem[Komissarov(2007)]{kom07} S. S. Komissarov 2007, \mnras, 382, 995
\bibitem[Lesch \& Birk(1998)]{lb98} H. Lesch and G. T. Birk 1998, ApJ, 499, 167
\bibitem[Lyubarsky(2003)]{lyu03} Y. Lyubarsky 2003, \mnras, 345, 153
\bibitem[Lyubarsky(2005)]{lyu05} Y. Lyubarsky 2005, \mnras, 358, 113
\bibitem[Lyubarsky \& Kirk(2001)]{lyu01} Y. Lyubarsky and J. G. Kirk 2001, ApJ, 547, 437
\bibitem[Lyubarsky \& Liverts(2008)]{lyu08} Y. Lyubarsky and M. Liverts 2008, \apj, 682, 1436
\bibitem[Lyutikov(2003)]{lyut03a} M. Lyutikov 2003, \mnras, 346, 540
\bibitem[Lyutikov(2006)]{lyut06} M. Lyutikov 2006, \mnras, 367, 1594
\bibitem[Lyutikov \& Uzdensky(2003)]{lyut03b} M. Lyutikov and D. {Uzdensky} 2003, \apj, 589, 893
\bibitem[Mart{\'{\i}} \& M{\"u}ller(2003)]{marti03} J.~M. Mart{\'{\i}} and E. M{\"u}ller 2003, {\itshape Living Reviews in Relativity}, 6, 7
\bibitem[Michel(1982)]{michel82} F. C. {Michel} 1982, Rev. Mod. Phys., 54, 1
\bibitem[Michel(1994)]{michel94} F. C. {Michel} 1994, \apj, 431, 397
\bibitem[Mignone et al.(2009)]{mig09} A. Mignone, M. Ugliano and G. Bodo, \mnras, 393, 1141
\bibitem[Mizuno et al.(2006)]{mizuno06} Y. Mizuno, K. I. Nishikawa, S. Koide, P. Hardee, and G. J. Fishman 2006, submitted to ApJS (astro-ph/0609004)
\bibitem[Noble et al.(2006)]{noble06} S.~C. {Noble}, C.~F. {Gammie},  J.~C. {McKinney} and L. {Del Zanna} 2006, \apj, 641, 626
\bibitem[Nunohiro \& Hirano(2003)]{nuno03} E. Nunohiro and S. Hirano 2003, {\itshape Transactions of the Japan Society for Industrial and Applied Mathematics}, 13, 159
\bibitem[Nunohiro et al.(1996)]{nuno96} E. Nunohiro, M. Suga and S. Hirano 1996, {\itshape Transactions of the Japan Society for Industrial and Applied Mathematics}, 6, 173
\bibitem[Parker(1957)]{parker} E. N. Parker 1957, \jgr, 62, 509
\bibitem[Petschek(1964)]{petschek} {H.~E. Petschek} 1964, ``Magnetic Field Annihilation'' in AAS/NASA Symposium on the Physics of Solar Flares, W. N. Ness, Ed. (NASA, Washington, DC, 1964), p. 425
\bibitem[Pritchett et al.(1996)]{prit96} P. L. Pritchett, F. V. Coroniti and V. K. Decyk 1996, \jgr, 101, 27413
\bibitem[Scholer(1989)]{scho89} M. Scholer 1989, \jgr, 94, 8805
\bibitem[Spitkovsky(2006)]{spit06} A. Spitkovsky 2006, \apj, 648, 51
\bibitem[Sweet(1958)]{sweet} P. A. Sweet 1958, in IAU Symp. 6, Electromagnetic Phenomena in Cosmical Physics, ed. B. Lehnert (New York: Cambridge Univ. Press), 123
\bibitem[Thompson \& Duncan(1995)]{thom95} C. Thompson and R. C. Duncan 1995, \mnras, 275, 255
\bibitem[Thompson \& Duncan(2001)]{thom01} C. Thompson and R. C. Duncan 2001, \apj, 561, 980
\bibitem[Ugai \& Tsuda(1977)]{ugai77} M. Ugai and T. Tsuda 1977, {\itshape J. Plasma Physics}, 17, 337
\bibitem[Watanabe \& Yokoyama(2006)]{naoyuki06} N. {Watanabe} and T. {Yokoyama} 2006, \apj, 647, L123 (WY06)
\bibitem[Woods \& Thompson(2006)]{woods06} P. M. Woods and C. Thompson 2006, in Compact stellar X-ray sources, ed. by W. Lewin \& M. van der Klis (Cambridge University Press), 547
\bibitem[Yokoyama \& Shibata(2001)]{yoko01} T. Yokoyama and S. Shibata 2001, \apj, 549, 1160
\bibitem[Zelenyi \& Krasnoselskikh(1979)]{zelenyi79} L. M. {Zelenyi} and V. V. {Krasnoselskikh} 1979, Astronomicheskii Zhurnal, 56, 819
\bibitem[Zenitani \& Hesse(2008a)]{zeni08b} S. Zenitani and M. Hesse 2008a, \pop, 15, 022101
\bibitem[Zenitani \& Hesse(2008b)]{zeni08c} S. Zenitani and M. Hesse 2008b, \apj, 684, 1477
\bibitem[Zenitani \& Hoshino(2001)]{zeni01} S. Zenitani and M. Hoshino 2001, \apj, 562, L63
\bibitem[Zenitani \& Hoshino(2005a)]{zeni05a} S. Zenitani and M. Hoshino 2005a, \apj, 618, L111
\bibitem[Zenitani \& Hoshino(2005b)]{zeni05b} S. Zenitani and M. Hoshino 2005b, \prl, 95, 095001
\bibitem[Zenitani \& Hoshino(2007)]{zeni07} S. Zenitani and M. Hoshino 2007, \apj, 670, 702
\bibitem[Zenitani \& Hoshino(2008)]{zeni08} S. Zenitani and M. Hoshino 2008, \apj, 677, 531
\end{thebibliography}
\end{document}